\documentclass[showpacs,twocolumn,preprintnumbers,nofootinbib,10pt]{revtex4-1}
\usepackage{graphicx}  
\usepackage{dcolumn}
\usepackage{bm}
\usepackage{amsmath,amssymb,amsfonts}
\usepackage{latexsym}
\usepackage{color}
\usepackage{subcaption}
\usepackage{ulem}

\usepackage[usestackEOL]{stackengine} 
 
\usepackage{multirow} 

\usepackage{hyperref}
\hypersetup{%
setpagesize=false,
 bookmarksnumbered=true,%
 bookmarksopen=true,%
 colorlinks=true,%
 linkcolor=blue,
 citecolor=magenta,
}

\usepackage{graphicx}
\usepackage{dcolumn}
\usepackage{bm}
\usepackage{amsmath,amssymb,amsfonts}
\usepackage{latexsym}
\usepackage{color}

\usepackage{subcaption}
\usepackage{ulem}

\usepackage{comment}

\def\nn    {\nonumber}

\def\gaa{{G_{\phi_1\phi_1}}}
\def\gbb{{G_{\phi_2\phi_2}}}
\def\gcc{{G_{\phi_3\phi_3}}}
\def\gdd{{G_{\phi_4\phi_4}}}

\def\tin{{t_{\rm{in}}}}
\def\tnd{{t_{\rm{end}}}}

\def\cchi{{c_\chi}}
\def\c2chi{{c_{2\chi}}}

\def\mpl{{M_{\rm{P}}}}

\begin{document}

\title{\boldmath The $R^2$-Higgs inflation with two Higgs doublets}

\author{Sung Mook Lee$^1$, Tanmoy Modak$^2$, Kin-ya Oda$^3$, Tomo Takahashi$^4$}
\affiliation{
$^1$Department of Physics \& IPAP \& Lab for Dark Universe, Yonsei University, Seoul 03722, Korea\\
$^2$Institut f{\"u}r Theoretische Physik, Universit{\"a}t Heidelberg, 69120 Heidelberg, Germany\\
$^3$Department of Mathematics, Tokyo Woman's Christian University, Tokyo 167-8585, Japan\\
$^4$Department of Physics, Saga University, Saga 840-8502, Japan}

\bigskip

\begin{abstract}
We study $R^2$-Higgs inflation in a model with two Higgs doublets in which  the Higgs sector
of the Standard Model is extended by an additional Higgs doublet, thereby four scalar fields are involved in the inflationary evolutions. 
We first derive the set of equations required to follow the
inflationary dynamics in this two Higgs doublet model, 
allowing a nonminimal coupling between the Higgs-squared and the Ricci scalar $R$, as well as the $R^2$ term
in the covariant formalism. By numerically solving the system of equations, we find that, in parameter space where a 
successful $R^2$-Higgs inflation are realized and consistent with low energy constraints, the inflationary dynamics can be 
effectively described by a single slow-roll formalism even though four fields are involved in the model.
We also argue that the parameter space favored by $R^2$-Higgs inflation
requires nearly degenerate masses for $m_\mathsf{H}$, $m_A$ and $m_{\mathsf{H}^\pm}$, where
$\mathsf{H}$, $A$, and $\mathsf{H}^\pm$ are the extra CP even, CP odd, and charged Higgs bosons
in the general two Higgs doublet model taking renormalization group
evolutions of the parameters into account. Discovery of such
heavy scalars at the Large Hadron Collider (LHC) are possible 
if they are in the sub-TeV mass range. Indirect evidences may also emerge 
at the LHCb and Belle-II experiments, however, to probe the quasi degenerate mass
spectra one would likely require high luminosity LHC or future lepton colliders such as the International Linear Collider and
the Future Circular Collider.
\end{abstract}

\maketitle

\section{Introduction}
The cosmic inflation~\cite{Starobinsky:1980te,Sato:1980yn,Guth:1980zm} 
can  successfully account for the observed flatness, horizon and the absence of the exotic-relics and, can seed 
the condition required for the subsequent hot big bang via the reheating process. 
The primordial density perturbations generated during inflation~\cite{Mukhanov:1981xt,Starobinsky:1982ee,Hawking:1982cz,Guth:1982ec}
can subsequently develop into large scale structure of the Universe and the cosmic microwave background (CMB) anisotropies measured by 
experiments such as Planck~\cite{Planck:2018jri}. While the cosmic inflation 
is indeed a well established paradigm for the very early epoch of the Universe, however, the mechanism 
behind it is still unknown. 

The Higgs inflation~\cite{Bezrukov:2007ep,Barvinsky:2008ia,Bezrukov:2010jz,Bezrukov:2013fka,DeSimone:2008ei,Bezrukov:2008ej,Barvinsky:2009ii} 
(for earlier works which employed essentially 
the same idea, see~\cite{Spokoiny:1984bd,Futamase:1987ua,Salopek:1988qh,Fakir:1990eg,Amendola:1990nn,Kaiser:1994vs,Cervantes-Cota:1995ehs,Komatsu:1999mt}) 
is one of the candidates that best fits
the CMB data~\cite{Planck:2018jri} and, draws significant attention due to its direct connection to the physics at the LHC. 
In the Standard Model (SM) Higgs inflation, the Higgs field $\Phi$ couples to 
the Ricci scalar $R$ via $\xi \Phi^\dagger \Phi R$ term, where $\xi $ is dimensionless
nonminimal coupling, and can account for the amplitude of the primordial perturbation along with the spectral index and the tensor-to-scalar ratio 
within the experimentally measured values~\cite{Planck:2018jri}. While the Higgs inflation 
can fit the CMB data without requiring any additional degrees of freedom between the electroweak and Planck
scale, however, a unitarity violating scale emerges below the Planck scale~\cite{Burgess:2009ea,Barbon:2009ya,Burgess:2010zq,Hertzberg:2010dc}.
Because the energy scale for inflation lies below such cut-off scale, it does not pose any problem for 
inflationary dynamics during the inflation  ~\cite{Bezrukov:2010jz}. However, during preheating stage i.e.,
when the inflaton field oscillates around the potential minima, longitudinal gauge bosons with momenta beyond the unitarity cut-off scale are produced 
violently~\cite{DeCross:2015uza,Ema:2016dny,Sfakianakis:2018lzf}. 
The perturbative unitarity of the Higgs inflation can be restored up to the Planck scale by
introducing additional scalars at the inflationary scale~\cite{Giudice:2010ka,Lebedev:2011aq} or,
by scalaron degree of freedom due to the presence of $R^2$ term ($R^2$-Higgs inflation) in the Jordan frame~\cite{Ema:2017rqn}
(see also for e.g.~\cite{Salvio:2015kka,Pi:2017gih,Gorbunov:2018llf,Gundhi:2018wyz,He:2018mgb,Cheong:2019vzl,Bezrukov:2019ylq,He:2020ivk,Bezrukov:2020txg,He:2020qcb}).

In this article we study the $R^2$-Higgs inflation in the general two Higgs doublet model (g2HDM) where
the SM is extended by an additional scalar doublet $\Phi'$. After the discovery of 125 GeV
Higgs boson $h$~\cite{ATLAS:2012yve,CMS:2012qbp} the existence of additional scalar doublet seems plausible
as all known fermions appear in nature with more than one generation. In addition, it is known that the electroweak 
vacuum is metastable for the current central values of the SM parameters~\cite{Degrassi:2012ry}, especially 
for top quark mass,  which also could pose a threat for the SM Higgs inflation~\footnote{If one demands the stability up to Planck scale, 
the required upper limit on the top quark pole mass is $m_t^\text{pole}\lesssim 171.4$\,GeV~\cite{Hamada:2014wna}, which is consistent 
at 1.6\,$\sigma$ with the current combined result $172.5\pm 0.7$\,GeV~\cite{ParticleDataGroup:2020ssz}.}.
If the scale of the instability is smaller than the required mass scale of the scalaron $ M \gtrsim 10^{-5}\mpl$ 
($\mpl\equiv 1/\sqrt{8\pi G}\simeq 2.4\times 10^{18}$ GeV) to fit the Planck measurement 
of the scalar power spectrum amplitude, just adding $R^2$ term may not be enough to solve the problem~\cite{Ema:2017rqn,He:2018gyf,Gorbunov:2018llf}. 
This partially motivates us to consider extension of the Higgs sector, in addition to the $R^2$ term of the SM Higgs inflation.

In this paper we study the inflationary dynamics and primordial fluctuations in 
the $R^2$-Higgs inflation in  the framework of the g2HDM based on 
the covariant formalism.  The work here also remedies the shortcomings of Ref.~\cite{Modak:2020fij} where inflationary dynamics was also under scrutiny due to
the unitarity violation by the required large nonminimal couplings $\mathcal{O}(10^4 - 6\times 10^4)$ as in the SM~\footnote{
See also Refs.~\cite{Gong:2012ri,Dubinin:2017irg,Choubey:2017hsq,Wang:2021ayg} for discussions on inflation in the 2HDM.}.
As we will argue, the parameter sets consistent with current observations of Planck and low energy constraints give almost
the same predictions for primordial power spectrum, we take four benchmark points as representative ones to show the inflationary dynamics and the evolutions of perturbations. 
For two benchmark points (PBs), we take the nonminimal coupling of the scalaron degree of freedom to be much larger than Higgs nonminimal coupling ($R^2$-like scenario) i.e., 
akin to the original Starobinsky model~\cite{Starobinsky:1979ty}, whereas for the other BPs, 
we take both the Higgs and scalaron nonminimal couplings relatively large (denoted as mixed $R^2$-Higgs scenario). 
We further provide sub-TeV parameter space for $R^2$-Higgs inflation in the g2HDM that can satisfy all 
observational constraints from Planck 2018~\cite{Planck:2018jri} and
discuss the possibility of probing such parameter space at the current experiment 
such as the LHC and future lepton colliders such as the International Linear Collider (ILC) and the Future Circular Collider (FCC-ee).
Moreover, indirect evidences of such additional Higgs bosons may also emerge in the ongoing flavor experiments such as LHCb and Belle-II.

The paper is organized as follows. In Sec.~\ref{form} we first discuss the model framework of
the g2HDM. We outline  the framework to follow the inflationary dynamics and perturbations based on the covariant formalism 
in Sec.~\ref{infdyna} followed by 
numerical study in Sec.~\ref{resul}. We discuss  possible discoveries and probes for the parameter space required
for $R^2$-Higgs inflation at the collider experiments in Sec.~\ref{coll}. We summarize our results with an outlook 
in Sec.~\ref{disc}.

\section{Model framework}\label{form}
The most general $CP$-conserving two Higgs doublet model~\footnote{See Refs.\cite{Djouadi:2005gj,Branco:2011iw} 
for pedagogical reviews on the two Higgs doublet model.} potential 
can be given in the Higgs basis as~\cite{Davidson:2005cw,Hou:2017hiw}
\begin{align}
& V(\Phi,\Phi') = \mu_{11}^2|\Phi|^2 + \mu_{22}^2|\Phi'|^2
            - (\mu_{12}^2\Phi^\dagger\Phi' + h.c.) +  \nn \\
 &\qquad \frac{\eta_1}{2}|\Phi|^4+\frac{\eta_2}{2}|\Phi'|^4
           + \eta_3|\Phi|^2|\Phi'|^2  + \eta_4 |\Phi^\dagger\Phi'|^2 + \nn\\
& \qquad \big[\frac{\eta_5}{2}(\Phi^\dagger\Phi')^2 
     + \left(\eta_6 |\Phi|^2 + \eta_7|\Phi'|^2\right) \Phi^\dagger\Phi' + \text{h.c.}\big],
\label{pot}
\end{align}
where the vacuum expectation value $v$ arises from the doublet $\Phi$ 
via the minimization condition $\mu_{11}^2=-\frac{1}{2}\eta_1 v^2$, 
while we take $\left\langle \Phi\right\rangle =(0,~v/\sqrt{2})^T$, $\left\langle \Phi'\right\rangle =0$ (hence $\mu_{22}^2 > 0$), 
and $\eta_i$s are quartic couplings. 
A second minimization condition, $\mu_{12}^2 = \frac{1}{2}\eta_6 v^2$, 
removes $\mu_{12}^2$, and 
the total number of parameters are reduced to nine.
The mixing angle $\gamma$ is given by, when diagonalizing the mass-squared matrix for $h$, $\mathsf{H}$,
\begin{align}
 c_\gamma^2 = \frac{\eta_1 v^2 - m_h^2}{m_\mathsf{H}^2-m_h^2},~\quad \quad \sin{2\gamma} = \frac{2\eta_6 v^2}{m_\mathsf{H}^2-m_h^2} \,,
 \label{mixin} 
\end{align}
with shorthand notation $c_\gamma= \cos\gamma$. The physical scalar masses can be expressed 
in terms of the parameters in Eq.~(\ref{pot}),
\begin{align}
 &m_{h,\mathsf{H}}^2 = \frac{1}{2}\bigg[m_A^2 + (\eta_1 + \eta_5) v^2 \mp \nn\\
 &\qquad\sqrt{\left(m_A^2+ (\eta_5 - \eta_1) v^2\right)^2 + 4 \eta_6^2 v^4}\bigg],\label{eq:mHh}\\
 &m_{A}^2 = \frac{1}{2}(\eta_3 + \eta_4 - \eta_5) v^2+ \mu_{22}^2,\label{eq:mA}\\
 &m_{\mathsf{H}^\pm}^2 = \frac{1}{2}\eta_3 v^2+ \mu_{22}^2.\label{eq:mHpm}
\end{align}

The scalars $h$, $\mathsf{H}$, $A$ and $\mathsf{H}^\pm$ couple to fermions by~\cite{Davidson:2005cw}
\begin{align}
\mathcal{L} = 
&-\frac{1}{\sqrt{2}} \sum_{F = U, D, L}
 \bar F_{i} \bigg[\big(-\lambda^F_{ij} s_\gamma + \rho^F_{ij} c_\gamma\big) h +\nn\\
 &\big(\lambda^F_{ij} c_\gamma 
 + \rho^F_{ij} s_\gamma\big)\mathsf{H} -i ~{\rm sgn}(Q_F) \rho^F_{ij} A\bigg]  P_R\; F_{j}\nn\\
 &-\bar{U}_i\left[(V\rho^D)_{ij} P_R-(\rho^{U\dagger}V)_{ij} P_L\right]D_j \mathsf{H}^+ - \nn\\
 &\bar{\nu}_i\rho^L_{ij} P_R \; L_j \mathsf{H}^+ +{\rm h.c.},\label{eff}
\end{align}
where $P_{L,R}\equiv (1\mp\gamma_5)/2$, $i,j = 1, 2, 3$ are generation indices, $V$ is  Cabibbo-Kobayashi-Maskawa matrix, $s_\gamma= \sin\gamma$ 
and $U=(u,c,t)$, $D = (d,s,b)$, $L=(e,\mu,\tau)$ and $\nu=(\nu_e,\nu_\mu,\nu_\tau)$ are vectors in flavor space. 
The matrices $\lambda^F_{ij}\; (=\sqrt{2}m_i^F/v)$ are real and diagonal,
whereas $\rho^F_{ij}$ are in general complex and non-diagonal. 

In general, one may allow data to constrain different elements of $\rho^F_{ij}$ matrices. However,
it is likely that $\rho^F_{ij}$ matrices follow the same flavor organization principle as in SM.
This means $\rho^F_{ij}\sim \lambda^F_{ii}$ i.e.,  
$\rho^{U}_{tt} \sim \lambda^U_t$, $\rho^{D}_{bb} \sim \lambda^D_b$, $\rho^{L}_{\tau\tau} \sim \lambda^L_\tau$ etc.
with suppressed off diagonal elements. While apart from getting involved in the RGE, the additional Yukawa 
couplings $\rho^F_{ij}$ do not play any major role in the inflationary dynamics, they are
essential for possible discovery of the heavy Higgs bosons $\mathsf{H}$, $A$ and $\mathsf{H}^\pm$. For all practical
purposes we shall set all $\rho^F_{ij}$ couplings to zero except for $\rho^{U}_{tt}$ and $\rho^{U}_{tc}$ throughout
this paper, however, the impact of turning on different $\rho^F_{ij}$ couplings and their constraints will be discussed
in Sec.~\ref{coll} of this paper. In this work, we primarily focus on the sub-TeV mass range i.e.  
$m_A$, $m_\mathsf{H}$, $m_{\mathsf{H}^\pm}$ in the range of $ 200 - 800$ GeV in the urge of finding complementarity between $R^2$-Higgs inflation and 
the ongoing collider experiments such as the LHC, although heavier Higgs bosons are also possible in principle.

\section{Inflationary Dynamics of $R^2$-Higgs inflation}\label{infdyna}
In this section, we outline the required formalism for $R^2$-Higgs inflation in the g2HDM and analyze perturbation 
theory using the covariant formalism \cite{Sasaki:1995aw,Kaiser:2010yu,Gong:2011uw,Peterson:2011yt,White:2012ya,Greenwood:2012aj,Kaiser:2013sna,Karamitsos:2017elm}.

\subsection{The action in $R^2$-Higgs inflation} \label{JordanEin:action}
The model can be understood as a particle-physics motivated generalization 
of $R^2$-Higgs inflation model. In the Jordan  frame, the action is given by 
\begin{align}
S & = \int d^4 x \sqrt{-g_J}\bigg[ -g_J^{\mu\nu}\left(\partial_\mu\Phi^\dagger \partial_\nu\Phi+
\partial_\mu\Phi'^\dagger \partial_\nu\Phi'\right) \nn\\
&+ \bigg(\frac{M_{P}^{2}}{2}+  \xi_{11} |\Phi|^2 + \xi_{22}
	|\Phi'|^2 + \left(\xi_{12} \Phi^\dagger \Phi'+ h.c.\right) \bigg)R_J  + \nn\\
	&\qquad\qquad\frac{\xi_R}{4} R_J^2 -V(\Phi,\Phi')\bigg] \,, \label{eq:jordan}
\end{align}
with $g_J = \det{g_{J\mu\nu}}$, Ricci scalar  $R_J$ and $(-1,+1,+1,+1)$ metric convention and we adopt the natural unit $\hbar=c=1$.
The $\xi_{ij}$s are nonminimal couplings between Higgs' and Ricci
Scalar and $\xi_R$ is the self coupling of Ricci scalar. In the following we would turn off the nonminimal 
coupling $\xi_{12} = \xi_{22} = 0$ for simplicity however we shall return to their impacts in the latter half of the paper.

We introduce an auxiliary field $s$ for which the action in Eq.~\eqref{eq:jordan}
can be rewritten as 
\begin{align}
	S = &\int d^4 x \sqrt{-g_J}\bigg[\left(\frac{M_P^2}{2}+  \xi_{11} \vert \Phi \vert^2  + \frac{1}{2}\xi_R s \right)  R_J - \frac{\xi_R}{4} s^2 \nn \\
	& -g_J^{\mu\nu}\left(\partial_\mu\Phi^\dagger \partial_\nu\Phi+ \partial_\mu\Phi'^\dagger
	\partial_\nu\Phi'\right)  -V(\Phi,\Phi')\bigg] \,, \label{eq:jordan1}
\end{align}
such that the variation of the action with respect to $s$ gives $s =R_J$.
For inflationary dynamics we choose the Higgs fields in the electromagnetic preserving direction:
\begin{align}
\Phi =
\frac{1}{\sqrt{2}} \begin{pmatrix}
  0 \\
  \rho_1 \\
\end{pmatrix} && \mbox{and}&&
\Phi' = \frac{1}{\sqrt{2}} 
 \begin{pmatrix}
   0\\
   \rho_2 +  i\rho_3\\
\end{pmatrix}. \label{filedparam}
\end{align}

We now perform the Weyl transformation to find the action in Einstein frame via
\begin{align}
 g_{\mu\nu}= F^2 g_{J\mu\nu} ,
\end{align}
where the conformal factor $F^2$ reads as
\begin{align}
 F^2 = 1+ \frac{\xi_{11} \rho_1^2 + \xi_R s}{M_P^2}.
\end{align}
The action of Eq.~\eqref{eq:jordan1} can be written in the Einstein frame as
\begin{align}
 S_E =& \int d^4 x \sqrt{-g} \bigg[\frac{M_P^2}{2}R- \frac{3 M_P^2}{4} (\partial_\mu \log( F^2))^2 \nn\\
 &-\frac{1}{2} \frac{(\partial_\mu \rho_1)^2+ (\partial_\mu \rho_2)^2 + (\partial_\mu \rho_3)^2}{F^2} -V_E\bigg],\label{eq:Ein}
\end{align}
where
\begin{align}
 V_E = \frac{V(\rho_1,\rho_2, \rho_3) + 2 \xi_R s^2}{8 F^4},\label{pot}
\end{align}
with
 \begin{align}
V(\rho_1,\rho_2,\rho_3)= &
 \bigg[\tilde\eta_1 \rho_1^4 + \tilde \eta_2 \left(\rho_2^2+\rho_3^2\right)^2 +2 \tilde \eta_5\left(\rho_2^2-\rho_3^2\right) \rho_1^2+\nn\\
 & \qquad  2\left(\tilde \eta_3+ \tilde \eta_4\right) \left(\rho_2^2+\rho_3^2\right) \rho_1^2+\nn\\
  & \qquad 4 \rho_2 \rho_1 \left\{ \tilde \eta_6 \rho_1^2 
   +\tilde \eta_7 \left(\rho_2^2+\rho_3^2\right) \right\} \bigg].
\label{potenexpn}
\end{align}
Here, $\tilde{\eta}_i$s correspond to the renormalization group evolution (RGE) of the 
parameters $\eta_i$ at the inflationary scale $ \sim O(H) $. Details of the running of the parameters are discussed in Section \ref{resul}.

Let us perform following field redefinition~\cite{Gong:2012ri}:
\begin{align}
\varphi= \sqrt{\frac{3}{2}} M_P \ln\left(F^2\right),
\end{align}
resulting in a simple form of the action
\begin{align}
 S_E = \int d^4 x \sqrt{-g} \bigg[\frac{R}{2}- 
 \frac{1}{2}G_{IJ}g^{\mu\nu}\partial_\mu{\phi_I} \partial_\nu\phi_J-V_E(\phi^I)\bigg],\label{action:Ein}
\end{align}
where $\phi^I =\{\varphi,\rho_1,\rho_2,\rho_3\}$ and $G_{IJ}$ is field space metric, with only non-vanishing components are diagonal:
\begin{align}
\gaa =1,~~\gbb = \gcc = \gdd =e^{-\sqrt{\frac{2}{3}}\frac{\varphi}{M_P}}.
\end{align}

Finally, we have the following action in the Einstein frame as
\begin{align}
 S_E =& \int d^4 x \sqrt{-g} \bigg[\frac{M_P^2}{2}R- \frac{1}{2} (\partial_\mu \varphi)^2 - \frac{1}{2} e^{-\sqrt{\frac{2}{3}}\frac{\varphi}{M_P}}\nn\\
 & \big((\partial_\mu \rho_1)^2+ (\partial_\mu \rho_2)^2 + 
 (\partial_\mu \rho_3)^2\big) -V_E\bigg],\label{eq:Ein}
\end{align}
with
\begin{align}
V_E(\varphi,\rho_1,\rho_2,\rho_3)&= \frac{1}{8}e^{-2\sqrt{\frac{2}{3}}\frac{\varphi}{M_P}}\bigg[ V(\rho_1,\rho_2,\rho_3) +2 \frac{M_P^4}{\xi_R}\nn\\
&\qquad \bigg(e^{\sqrt{\frac{2}{3}}\frac{\varphi}{M_P}} - 1-\frac{\xi_{11}}{M_P^2}\rho_1^2\bigg)^2\bigg] \,.  \label{pot:inf}
\end{align}
During numerical analysis, to remain in the perturbative regime, we also demand the upper bound on the scalaron mass 
as discussed in Refs.~\cite{Ema:2017rqn,Gorbunov:2018llf,He:2018gyf}.

The equation of motions for the fields $\phi^I$ can also be found by varying the action in 
Eq.~\eqref{action:Ein} with respect to $\phi^I$ as
\begin{align}
g^{\mu\nu} \partial_\nu\partial_\mu\phi^I + g^{\mu\nu} \Gamma^I_{JK} \partial_\mu \phi^J\partial_\nu \phi^K- G^{IK} V_{E,K}=0,\label{fildeq}
\end{align}
where $\Gamma^I_{JK}(\phi^{M})$ is the Christoffel symbol for the field space manifold  $G^{IK}$ and $V_{E,K}$ denotes
derivative of $V_E$ with respect to field $\phi^K$.
Explicit elements of $\Gamma^I_{JK}$ in our model are given in the Appendix~\ref{fieldchris}. 
The background dynamics is governed by the Friedmann equations: 
\begin{align}
  H^2 &= \frac{1}{3 \mpl^2} \left(\frac{1}{2} G_{IJ} \dot{\phi}^I \dot{\phi}^J + V_{E}(\phi^I)\right),\label{hubble1}\\ 
 \dot{ H}&= -\frac{1}{2 \mpl^2} G_{IJ} \dot{\phi}^I \dot{\phi}^J \,, \label{hubble2}
\end{align}
where an overdot represents the derivative with respect to time.

\subsection{Background Dynamics and the Perturbation Theory: Covariant Formalism} \label{App:Perturbation Theory}
In this section we outline the covariant
formalism~\cite{Sasaki:1995aw,Kaiser:2010yu,Gong:2011uw,Peterson:2011yt,White:2012ya,Greenwood:2012aj,Kaiser:2013sna,Karamitsos:2017elm,Kaiser:2012ak} 
for our inflationary model, which includes four scalar fields $\phi^I =\{\varphi,\rho_1,\rho_2,\rho_3\}$.
We closely follow the  
formalism for multi-field inflation as discussed in Ref.~\cite{Kaiser:2012ak}.
We divide the fields
into classical background part ($ \bar{\varphi}^I $) and perturbation part ($ \delta\phi^I $) as
\begin{align}
\phi^I(x^\mu) = \bar{\phi}^I(t) + \delta\phi^I(x^\mu)\label{fieldexpan}.
\end{align}

The perturbed spatially flat
Friedmann-Robertson-Walker (FRW) metric can be expanded as~\cite{Kodama:1984ziu,Mukhanov:1990me,Malik:2008im}
\begin{align}
ds^2 &= -(1+2A) dt^2 +2 a(t) (\partial_i B) dx^i dt +\nn\\
&a(t)^2 \left[(1-2\psi) \delta_{ij}+ 2 \partial_i \partial_j E\right] dx^i dx^j,
\end{align}
where $a(t)$ is scale factor and $t$ is the cosmic time. $A, B, \psi$ and $E$ characterize the scalar metric perturbations.

The $\phi^I(x^\mu)$ field value in Eq.~\eqref{fieldexpan} depends on the background field value $\bar{\phi}^I(t)$ 
and, gauge dependent field fluctuation $\delta\phi^I(x^\mu)$. This motivates one
to consider gauge independent Mukhanov-Sasaki variables for the field
fluctuations expressed as~\cite{Sasaki:1986hm,Mukhanov:1988jd,Mukhanov:1990me}
\begin{align}
Q^I = \mathcal{Q}^I + \frac{\dot{\bar{\phi}}^I}{H}\psi,
\end{align}
with $D_\kappa\phi^I\vert_{\kappa=0}= \frac{d\phi^I}{d\kappa}\vert_{\kappa=0} \equiv \mathcal{Q}^I$~\cite{Gong:2011uw}, where
$\kappa$ is the trajectory in the field space. 
The field fluctuations $\delta\phi^I$ 
can be expressed in series of $\mathcal{Q}^I$~\cite{Gong:2011uw,Elliston:2012ab} as 
\begin{align}
\delta\phi^I &= \mathcal{Q}^I -\frac{1}{2} \Gamma^I_{JK} \mathcal{Q}^I \mathcal{Q}^J+\frac{1}{3!} \big(\Gamma^I_{MN} \Gamma^N_{JK}-\Gamma^I_{JK,M}\big)\nn\\ 
&\qquad\qquad \qquad\qquad  \qquad\qquad  \mathcal{Q}^I \mathcal{Q}^J  \mathcal{Q}^M+\dots,
\end{align}
with Christoffel symbols $\Gamma^I_{JK}$ evaluated with background field. 
We remark that, while $\bar{\phi}^I$ are not vectors in the field-space manifold, $\mathcal{Q}^I$, $\dot{\bar{\phi}}^I$ and
$Q^I$ all transform as vectors in the field-space manifold. At this point it is useful to define the covariant derivative of vectors $S^I$ and $S_I$ in
the field-space as 
\begin{align}
\mathcal{D}_J S^I \equiv \partial_J S^I + \Gamma^I_{JK} S^K, 
\qquad
\mathcal{D}_J S_I \equiv \partial_J S_I - \Gamma^K_{IJ} S_K.
\end{align}
One can also define covariant derivative with respect to cosmic time $t$ 
as~\cite{Easther:2005nh,Langlois:2008mn,Peterson:2010np,Peterson:2010mv,Peterson:2011yt}
\begin{align}
\mathcal{D}_t S^I \equiv \dot{\bar{\phi}}^J \mathcal{D}_J S^I = \dot{S}^I + \Gamma^I_{JK} S^J \dot{\bar{\phi}}^K. 
\end{align}
With these definitions, one can find that the background field equations can be written as
\begin{align}
\mathcal{D}_t \dot{\bar{\phi}}^I + 3 H \dot{\bar{\phi}}^I + G^{IK} V_{E,K} =0.\label{eq:bkg}
\end{align}
Numerically we solve these set of background equations of motion for four fields $\{\varphi,\rho_1,\rho_2,\rho_3\}$
along with Eq.~\eqref{hubble1}. While solving
these equations we always check that the $\dot{ H}^2$ estimated from these solutions and, directly from the Eq.~\eqref{hubble2} 
are equal with high precision. Here, we remark that both in Eq.~\eqref{hubble1} and \eqref{hubble2}, all field
dependent quantities are evaluated with the background ones.

On the other hand, the equations for gauge invariant field fluctuations $Q^I$ are given by
\begin{align}
&\mathcal{D}^2_t Q^I + 3 H \mathcal{D}_t Q^I +\bigg(\frac{k^2}{a^2}\delta^I_J + \mathcal{M}^I_J- \frac{1}{\mpl^2 a^3} \nn\\
&\qquad\qquad\qquad\qquad\mathcal{D}_t\bigg(\frac{a^3}{H}\dot{\bar{\phi}}^I \dot{\bar{\phi}}_J\bigg) \bigg) Q^J=0,\label{eq:fluc}
\end{align}
where
\begin{align}
\mathcal{M}^I_J = G^{IM} \mathcal{D}_J\mathcal{D}_M V_E - R^I_{MNJ} \dot{\bar{\phi}}^M \dot{\bar{\phi}}^N\label{eq:masssqmat}
\end{align}
with $R^I_{MNJ}$ being field-space Riemann tensor, and we denote
\begin{align}
	\delta \mathcal{M}^{I}_{J} = - 
	\frac{1}{\mpl^2 a^3} \mathcal{D}_t\bigg(\frac{a^3}{H}\dot{\bar{\phi}}^I \dot{\bar{\phi}}_J\bigg)
\end{align}
for future use. Here in both Eq.~\eqref{eq:bkg} and Eq.~\eqref{eq:fluc} quantities such as $G^{IK}$, $\Gamma^I_{JK}$, $V_E$ etc.
all are evaluated with the background quantities.

One can re-express Eq.~\eqref{hubble1} and Eq.~\eqref{hubble2} as
\begin{align}
H^{2} &= \frac{1}{3 \mpl^2} \left(\frac{1}{2} \dot{\sigma}^2 + V_{E}\right),\label{hubblesigma1}\\ 
\dot{H}&= -\frac{1}{2 \mpl^2} \dot{\sigma}^2, \label{hubblesigma2}
\end{align}
where $\dot{\sigma}$ is the length of the velocity vector $\dot{\bar{\phi}}^I$ in field-space defined as
\begin{align}
\dot{\sigma} = \sqrt{G_{IJ} \dot{\bar{\phi}}^I \dot{\bar{\phi}}^J} \,.
\end{align}
We also introduce a unit vector $\hat{\sigma}^I$ given as
\begin{align}
\hat{\sigma}^I =\frac{\dot{\bar{\phi}}^I}{\dot{\sigma}}. 
\end{align}
The equation of motion reads as 
\begin{align}
\ddot{\sigma} + 3H \dot{\sigma} + V_{E,\sigma} = 0,\label{eqsigma}
\end{align}
where $V_{E,\sigma} \equiv \hat{\sigma}^I V_{E,I}$.
Together with Eqs.~\eqref{hubblesigma1}~and~\eqref{hubblesigma2}, Eq.~\eqref{eqsigma}
simply conforms of a single-field model with canonically normalized kinetic term.
The slow-roll parameters $\epsilon$ and $\eta_{\sigma\sigma}$ can be defined as
\begin{align}
&\epsilon \equiv -\frac{\dot{H}}{H^2}=\frac{3\dot{\sigma}^2}{\dot{\sigma}^2+2V_E},\label{epsi}\\
&\eta_{\sigma\sigma} \equiv \mpl^2 \frac{\mathcal{M}_{\sigma\sigma}}{V_E}, 
\end{align}
where $\mathcal{M}_{\sigma\sigma} \equiv \hat{\sigma}_I \hat{\sigma}^J \mathcal{M}^I_J = \hat{\sigma}^I \hat{\sigma}^J (\mathcal{D}_I\mathcal{D}_J V_E)$.
The energy density $\varrho(t)$ and pressure $p(t)$ of the scalar field multiplets can be written as 
\begin{align}
&\varrho = \frac{1}{2}\dot{\sigma}^2 + V_E,\label{energyden}\\
&p= \frac{1}{2}\dot{\sigma}^2 - V_E\label{pres}.
\end{align}

The field space directions orthogonal to $\hat{\sigma}^I$ are expressed as
\begin{align}
\hat{s}^{IJ} = G^{IJ}-\hat{\sigma}^I \hat{\sigma}^J.
\end{align}
The  $\hat{\sigma}^I$ and $\hat{s}^{IJ}$ vectors are related by the relations
\begin{align}
&\hat{\sigma}^I \hat{\sigma}_I = 1,\nn\\
&\hat{s}^{IJ}\hat{s}_{IJ} = N-1,\label{ortho}\\
&\hat{\sigma}_I \hat{s}^{IJ} = 0~\rm{for~each}~J\nn,
\end{align}
where $N$ is the number of scalar fields which is four in our case.

One can now decompose the perturbations in the directions of $\hat{\sigma}^I$ and $\hat{s}^{IJ}$ as
\begin{align}
&Q_\sigma = \hat{\sigma}_I Q^I \label{eq:Qsig},\\
&\delta s^I  =  \hat{s}^I_J Q^J\label{eq:deltas},
\end{align}
where $Q_\sigma$ and $\delta s^I$ are respectively called adiabatic and entropy perturbations.

In our four field case, there are three independent $\delta s^I$s.
It is convenient to define three additional unit vectors 
by which one can identify these independent entropy directions. Here we follow the 
decomposition as discussed in Ref.~\cite{Kaiser:2012ak} which essentially can reproduce
the kinematical basis of Refs.~\cite{Peterson:2010np,Peterson:2010mv,Peterson:2011yt}.
In this regard, we first define turning vector $\omega^I$ which can be defined as the covariant rate of change of $\hat{\sigma}^I$ i.e.,
\begin{align}
\omega^I =  \mathcal{D}_t  \hat{\sigma}^I.
\end{align}
It is also clear that with the definition above the turning vector is orthogonal to the $\hat{\sigma}^I$ i.e.
$\omega_I\hat{\sigma}^I=0$. The unit turning vector is defined as
\begin{align}
\hat{\omega}^I =\frac{\omega^I}{\omega},
\end{align}
with $\omega= |\omega^I|=\sqrt{G_{IJ} \omega^I\omega^J}$.
We now can construct a new projection operator $\gamma^{IJ}$
\begin{align}
\gamma^{IJ} = G^{IJ} -\hat{\sigma}^I \hat{\sigma}^J- \hat{\omega}^I \hat{\omega}^J.
\end{align}
Next vector is defined as
\begin{align}
\Pi^I = \frac{1}{\omega}\mathcal{M}_{\sigma J} \gamma^{IJ} \,,
\end{align}
with $\mathcal{M}_{\sigma J}= \hat{\sigma}_I\mathcal{M}^I_J$.
$\Pi^I$ is orthogonal to both $\hat{\sigma}_I$ and $\hat{\omega}_I$.
The corresponding unit vector can be defined as $\hat{\pi}^I = \Pi^I/\Pi$ with $\Pi=|\Pi^I|$
and, a projection operator defined as $q^{IJ} = \gamma^{IJ}-\hat{\pi}^I \hat{\pi}^J$. 
The final vector for our four field scenario is $\tau^I$ which is defined as
\begin{align}
\tau^I = \frac{1}{\Pi}\left(\mathcal{M}_{sJ} + \frac{\dot{\sigma}}{\omega} \hat{\sigma}^K \hat{\sigma}_N (\mathcal{D}_K \mathcal{M}^N_J)\right) q^{IJ} \,,
\end{align}
and its corresponding unit vector is $\hat{\tau}^I = \tau^I/|\tau^I|$.
With the unit vectors $\hat{\omega}^I$, $\hat{\pi}^I$ and $\hat{\tau}^I$,
we now are ready to define three independent components of entropy perturbations as
\begin{align}
&Q_s = \hat{\omega}_I Q^I,\\
&Q_u = \hat{\pi}_I Q^I,\\
&Q_v = \hat{\tau}_I Q^I.
\end{align}

The gauge-invariant curvature perturbation $\mathcal{R}$ is defined as~\cite{Mukhanov:1990me,Malik:2008im}
\begin{align}
\mathcal{R} = \psi -\frac{ H}{\varrho+p} \delta q,\label{gaugeinvR}
\end{align}
where $\varrho$ and $p$ are defined in Eq.~\eqref{energyden} and Eq.~\eqref{pres} and, $\delta q$
is the energy density flux defined by $T^0_i \equiv \partial_i \delta q$.
Utilizing 
\begin{align}
\delta q = - G_{IJ} \dot{\bar{\phi}}^I \delta\phi^J = -\dot{\sigma}\hat{\sigma}_I \delta\sigma^I,
\end{align}
and, Eqs.~\eqref{fieldexpan} and \eqref{eq:Qsig} we find that $\mathcal{R}$ can be given by
\begin{align}
\mathcal{R} = \frac{ H}{\dot{\sigma}} Q_\sigma\label{eq:curvpurt}.
\end{align}
The normalized entropy perturbations~\cite{Wands:2000dp,Amendola:2001ni,Wands:2002bn,Kaiser:2012ak} can be derived as
\begin{align}
\mathcal{S} = \frac{ H}{\dot{\sigma}} Q_s,\label{eq:entropy1}\\
\mathcal{U} = \frac{ H}{\dot{\sigma}} Q_u,\label{eq:entropy2}\\
\mathcal{V} = \frac{ H}{\dot{\sigma}} Q_v.\label{eq:entropy3}
\end{align}
At this point we remark that in our numerical analysis we always check that the orthogonality conditions of Eq.~\eqref{ortho}
and, as well as for the other unit vectors $\hat{\omega}^I$, $\hat{\pi}^I$ and $\hat{\tau}^I$ as given
in Ref.~\cite{Kaiser:2012ak} are satisfied.

Our focus of interest is the power spectrum of the gauge invariant curvature perturbation defined as~\cite{Mukhanov:1990me,Bassett:2005xm}
\begin{align}
\langle\mathcal{R}(\bm{k}_1) \mathcal{R}(\bm{k}_2) \rangle= (2\pi)^3 \delta^{(3)}(\bm{k}_1+\bm{k}_2) P_{\mathcal{R}}(k_1)
\end{align}
and $P_{\mathcal{R}}(k)= |\mathcal{R}|^2$. The dimensionless power spectrum for the adiabatic perturbation is given by
\begin{align}
\mathcal{P}_{\mathcal{R}}(t;k)= \frac{k^3}{2\pi^2}P_\mathcal{R}(k)\label{eq:powadia}. 
\end{align}
Similarly the power spectrum for the entropy perturbations are expressed as
\begin{align}
&\mathcal{P}_{\mathcal{S}}(t;k)= \frac{k^3}{2\pi^2}|\mathcal{S}|^2\label{eq:powentrop1},\\ 
&\mathcal{P}_{\mathcal{U}}(t;k)= \frac{k^3}{2\pi^2}|\mathcal{U}|^2\label{eq:powentrop2},\\
&\mathcal{P}_{\mathcal{V}}(t;k)= \frac{k^3}{2\pi^2}|\mathcal{V}|^2\label{eq:powentrop3}.
\end{align}
In order to find the power spectrum of the adiabatic and entropy perturbations given in Eqs.~\eqref{eq:powadia}, 
\eqref{eq:powentrop1}, \eqref{eq:powentrop2}, and \eqref{eq:powentrop3},  we utilize the quantities
$ H$, $\epsilon$ and unit vectors such as $\hat{\sigma}^I$, $\hat{\omega}^I$ etc. from the solutions of the 
Eqs.~\eqref{hubble1} and \eqref{eq:bkg} while $Q_\sigma$, $Q_s$, $Q_u$ and $Q_v$ are evaluated using the solutions of 
mode equations from Eqs.~\eqref{eq:fluc}. For a given Fourier mode $k$, we calculate the different power spectra at
the $t=\tnd$  numerically as a function of $k$ as
\begin{align}
&\mathcal{P}_{\mathcal{R}}(k)=\mathcal{P}_{\mathcal{R}}(\tnd;k),\label{eq:PR}\\
&\mathcal{P}_{\mathcal{S}}(k)=\mathcal{P}_{\mathcal{S}}(\tnd;k),\label{eq:PS}\\
&\mathcal{P}_{\mathcal{U}}(k)=\mathcal{P}_{\mathcal{U}}(\tnd;k),\label{eq:PU}\\
&\mathcal{P}_{\mathcal{V}}(k)=\mathcal{P}_{\mathcal{V}}(\tnd;k),\label{eq:PV}
\end{align}
where $\tnd$ denotes the time when inflation ends i.e. when $\epsilon =1$.

The spectral index $n_{s}$ of the power spectrum of the adiabatic fluctuations is defined as
\begin{align}
n_{s} = 1 + \frac{d\ln\mathcal{P}_{\mathcal{R}}(k)}{d\ln k}\label{specin1}. 
\end{align}
As will be discussed in the next section, although four fields are involved during inflation in our model, 
we argue that in the parameter space where  Planck and low energy constraints 
are satisfied, the power spectrum can effectively be described by the single field-like inflation. In such a case,
the spectral index can be calculated as
\begin{align}
n_{s}(t_\ast) \approx 1 -6 \epsilon(t_\ast) + 2 \eta_{\sigma\sigma}(t_\ast) \label{specin2},
\end{align}
where $t_\ast$ denotes the time when the reference scale exited the horizon
and the tensor-to-scalar ratio is given by $r=16 \epsilon$.

\begin{table*}[htbp]
		\begin{tabular}{|c |c| c| c| c | c | c| c | c |c| c| c| c}
			\hline
			BPs  &$\tilde\eta_1$ &  $\tilde\eta_2$   &  $\tilde \eta_3$   & $\tilde\eta_4$  & $\tilde\eta_5$ & $\tilde\eta_6$ & $\tilde\eta_7$  & $\xi_{11}$  & $\xi_{R}$\\
			\hline
			$a$         & 0.72459  & 0.834059 & -0.287252  & 0.489654    & -0.010900   & -0.510739 & 0.333532  & 1    & $2.4\times 10^9$\\
			$b$         & 0.845674 & 1.281688 & 0.017365   & 0.611085    & -0.776203   & -0.361704 & 0.050345  & 1800 & $2.25\times 10^9$\\
			$c$         & 2.08746  & 1.11479  & 2.56305    & -1.93179    & -0.0412796  & -0.521398 & -0.0743505 & $10^{-3}$ &  $2.42\times 10^9$ \\
			$d$         & 0.634249 & 2.98825  & 0.083228   & 0.087188    & 0.152301    & -0.494063 & 0.679174   & 200  &  $2.4\times 10^9$ \\
			\hline
			\hline
	\end{tabular}
	\caption{Benchmark points chosen for our analysis. See text for details.}
	\label{highscale}
\end{table*}

\begin{table*}[htbp]
	\centering
	\resizebox{\textwidth}{!}{	
		\begin{tabular}{|c |c| c| c| c | c | c| c | c |c| c| c| c|}
			\hline
			BPs     &   $\eta_1$ &  $\eta_2$   &  $\eta_3$   & $\eta_4$  & $\eta_5$ & $\eta_6$ & $\eta_7$  
			& $m_{H^+}$  & $m_A$ & $m_\mathsf{H}$ &  $\frac{\mu_{22}^2}{v^2}$ & $c_\gamma$\\ 
			& &&&&&&&& (GeV) & (GeV) & (GeV)&\\ 
			\hline
			$a$   &  0.258353 & 0.214212 & -0.104774  & 0.321234 & -0.00339535 & -0.0474321 & 0.132779     & 424  & 436 & 435  & 3.017 & 0.0165\\
			$b$   &  0.257981 & 0.363637 & -0.026754  & 0.194828 & -0.225193  & -0.0337426  & 0.0418302   & 429  & 443 & 428  & 3.043 & 0.0122 \\
			$c$   &  0.259349 & 0.245545 & 0.469357   & -0.579992 & -0.014849 & -0.050576  & 0.061887     & 347  & 322 & 321  & 1.756 &0.0352  \\
			$d$   &  0.258161 & 0.40482  & 0.134086   & 0.028604  & 0.059236  & -0.066433  & 0.086559     & 681  & 681 & 683  & 7.581  & 0.0089 \\
			\hline
			\hline
	\end{tabular}}
	\caption{The low energy parameters for the BPs shown in Table~\ref{highscale} along with the masses of heavy Higgs bosons and mixing angle $c_\gamma$
		between the CP even Higgs $h$ and $\mathsf{H}$.}
	\label{lowscale}
\end{table*}

To solve field fluctuations given in Eq.~\eqref{eq:fluc} we utilize the Bunch-Davies vacuum  
\begin{align}
Q^I (k\tau_c \longrightarrow -\infty) = \sqrt{\frac{1}{2k}} e^{-i k\tau_c}.
\end{align} 
Here $\tau_c$ is conformal time related to cosmological time $t$ via $dt = a d\tau_c$.
Such exact initial conditions need to be imposed in the infinite past which is numerically impractical.
Here we utilize the approximate initialization of the field fluctuations and impose them in the sufficiently 
past such that the Hubble parameter at that time remains approximately 
constant. The conditions is~\cite{Antusch:2015nla}
\begin{align}
&Q^I(\tin)  \simeq \frac{H}{\sqrt{2 k^3}}\bigg(i+ \frac{k}{a H}\bigg)e^{i\frac{k}{a H}}\label{inifluc1},
\end{align}
where $k$ is the corresponding Fourier mode~\footnote{For each mode, we initialize $Q^I$ about $5$ $e$-foldings before they exit horizon.}.
One can also use the approximate initial condition as in Ref.~\cite{Powell:2007gu}
\begin{align}
&Q^I(\tin)  \simeq \sqrt{\frac{1}{2k}} e^{-i \frac{k}{a H} \zeta},\label{inifluc2}
\end{align}
where $\zeta$ is a numerical prefactor which we assumed to be
$100$ to ensure the field fluctuations are initialized sufficiently
early times i.e. well within sub-horizon scale for each $k$ mode.
We have also checked numerically that both these initial conditions as in Eqs.~\eqref{inifluc1} and \eqref{inifluc2}
give the same power spectrum.

\section{Numerical results}\label{resul}
\subsection{Benchmark Parameters and RG Running}

\begin{figure}[htbp!]
\centering
\includegraphics[width=.4 \textwidth]{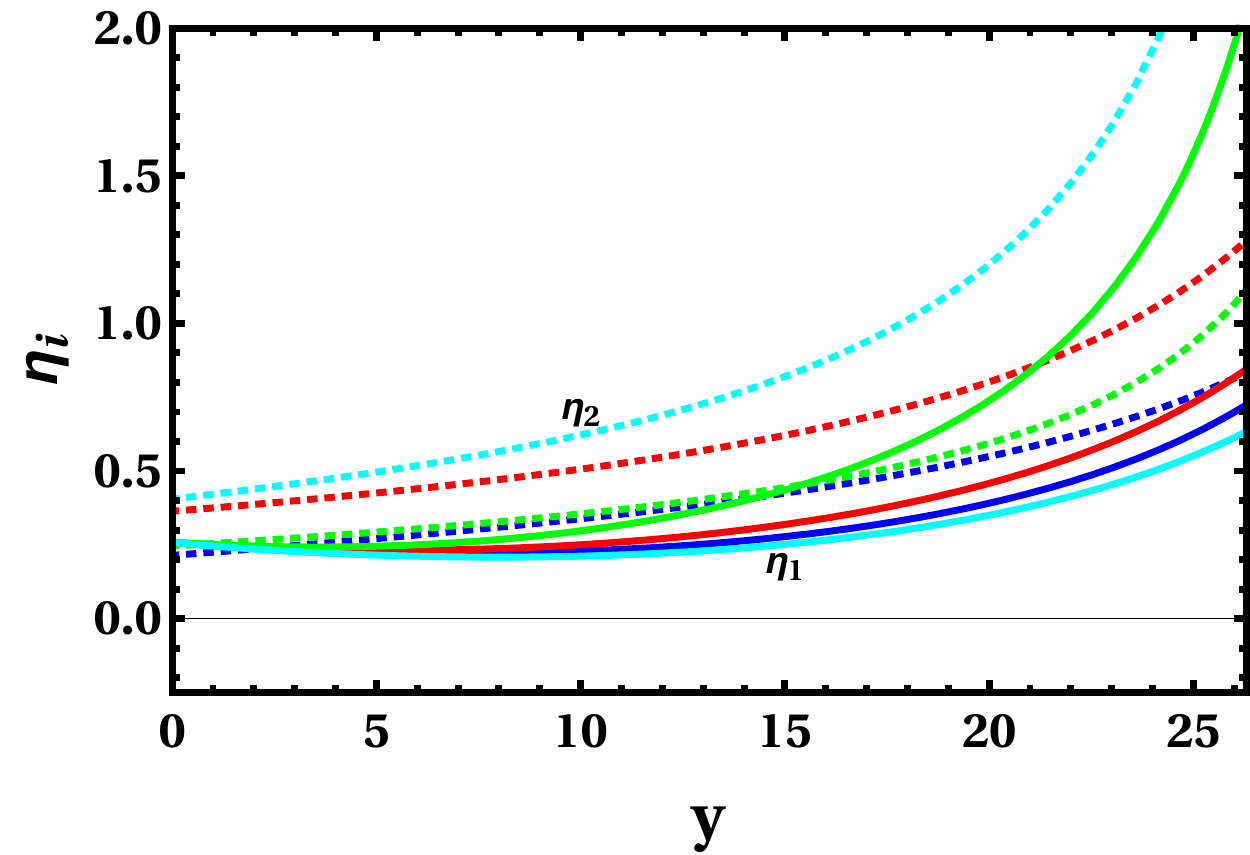}
\caption{The running of $\eta_1$ (solid) and $\eta_2$ (dotted) for BP$a$, $b$, $c$ and $d$ are shown in blue, red, green and cyan lines respectively.} 
\label{etapl}
\end{figure}

\begin{figure*}[htbp]
	\begin{subfigure}{.42\textwidth}
		\centering
		\includegraphics[width=.82\linewidth]{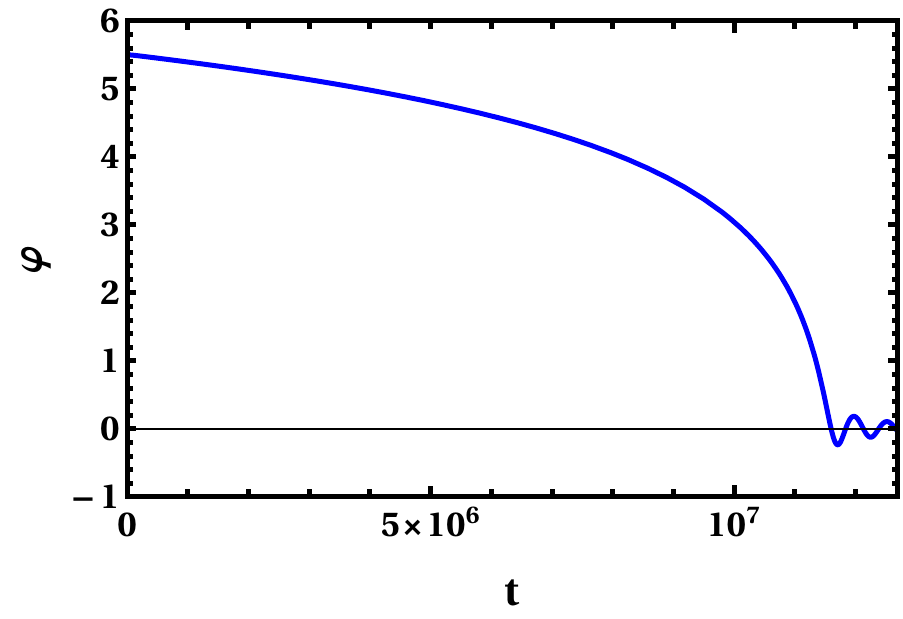}
	\end{subfigure}%
	\begin{subfigure}{.43\textwidth}
		\centering
		\includegraphics[width=.84\linewidth]{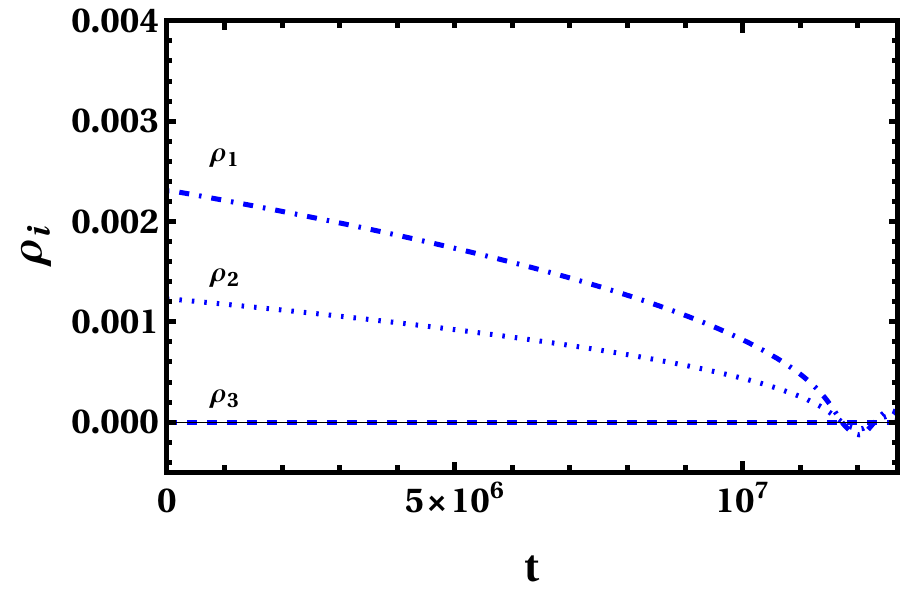}
	\end{subfigure}
	%
	%
	\caption{
%
		The time evolution of each background field $\bar{\phi}^I$ in $\mpl$ units for the BP$a$. The solid line in left panel correspond 
		to the evolution of $\varphi$ field while dot-dashed,
		dotted and dashed lines in right panels correspond to the evolution for $\rho_1$, $\rho_2$ and $\rho_3$ fields respectively.
		To find the background evolution we have used initial field values
		$\varphi^1(\tin)=5.5$, $\varphi^2(\tin)=0.002306$,
		$\varphi^3(\tin)=0.001227$ and $\varphi^4(\tin)=0$. 
		The prescription for finding the initial conditions are discussed in Appendix.~\ref{apprxini} 
		and the analytic relations between $ \varphi $ and $ \rho_{i} $ are given in Appendix.~\ref{Appendix:Valley}}
	\label{fieldevoR}
\end{figure*}

We consider four BPs for illustrative purpose which are  shown in Table~\ref{highscale}.  The BP$a$ and $c$ correspond to scenarios where the 
Higgs nonminimal coupling $\xi_{11}$ is small (denoted as $R^2$-like scenario). In BP$b$ and BP$d$ we consider parameter 
space  where $\xi_{11}$ is relatively
large (denoted as mixed Higgs-$R^2$ like scenario). 
The low energy values for the corresponding parameters in Table~\ref{highscale} are presented at low scale ($y=0$) in Table~\ref{lowscale}~\footnote{
Here we provide the values of the quartic coupling $\tilde{\eta}_i$ in Table~\ref{highscale} up to six
decimal place. As one should expect, we remark that $\tilde{\eta}_i$ values are 
highly sensitive to the corresponding low scale values of $\eta_i$ as given in Table~\ref{lowscale}, for which we also consider six decimal place.}.

We require the dynamical parameters in Eq.~\eqref{pot} to satisfy the  
unitarity, perturbativity, and positivity constraints at the low scale ($\mu = m_W$) for which we utilized 2HDMC~\cite{Eriksson:2009ws}.  
To match the convention of 2HDMC, we take $-\pi/2\leq \gamma \leq \pi/2$. For more details 
on the convention, parameter counting and low energy scanning we redirect readers to Refs.~\cite{Hou:2019qqi,Modak:2019nzl,Hou:2019mve,Modak:2020uyq}.
The low energy parameter sets for all 
BPs are further checked to satisfy the electroweak precision observables~\cite{Peskin:1991sw}
within the $2\sigma$ error~\cite{Baak:2014ora}. While they do not directly
play significant role in inflationary dynamics, we assumed $\rho^U_{tt} = 0.5$, $\rho^U_{tc}=0.2$, $\lambda^U_t=\sqrt{2}\frac{m_t}{v}$ at low scale
and set all other Yukawa couplings to zero for simplicity for RG running.

It has been found that for parameter sets where $|\eta_i| > 1$ at the low scale get 
generally excluded after imposing perturbativity criteria at the high scale \cite{Modak:2020fij}. 
Therefore, we simply adopt the strategy as in Ref.~\cite{Modak:2020fij} and considered
all benchmark points such that at low scale all $|\eta_{i}|$s are $\leq 1$.

For the RGE of the parameters in Eq.~\eqref{pot} as
well as the Yukawa couplings $\rho^F$ and $\lambda^F$ in the Eq.~\eqref{eff} we utilized the 
$\beta_x$ functions ($\beta_x \equiv \partial x /\partial y$ with $y \equiv \ln (\mu/ m_W)$ where $ \mu $ is the renormalization scale) for g2HDM
given in  Ref.~\cite{Ferreira:2015rha,Haber:1993an}. Here we take the low scale as $y=0$ and, 
take $y \approx 26$ as inflationary scale or high scale~\footnote{To be precise in
our numerical analysis we performed the RG evolution from $y=0$ to $y=26.3$.}.
After finding the parameters satisfying the constraints such as unitarity, 
perturbativity, stability and electroweak precision observables at the low (EW) scale, the 
same parameters are then evolved from low scale to high scale via the RG equations.

At the high scale, we also demand $|\tilde{\eta_i}|$ and the Yukawa couplings to be within $[-\pi,\pi]$. 
To ensure the positivity of the potential in Eq.~\eqref{potenexpn}
the quartic couplings $\tilde{\eta}_{1,2}$ are required to be positive, 
which is true for all four BPs as is evident from Fig~.\ref{etapl}.

\subsection{Background Dynamics and Power Spectrum}

\begin{figure*}
	\begin{subfigure}{.43\textwidth}
		\centering
		\includegraphics[width=.87\linewidth]{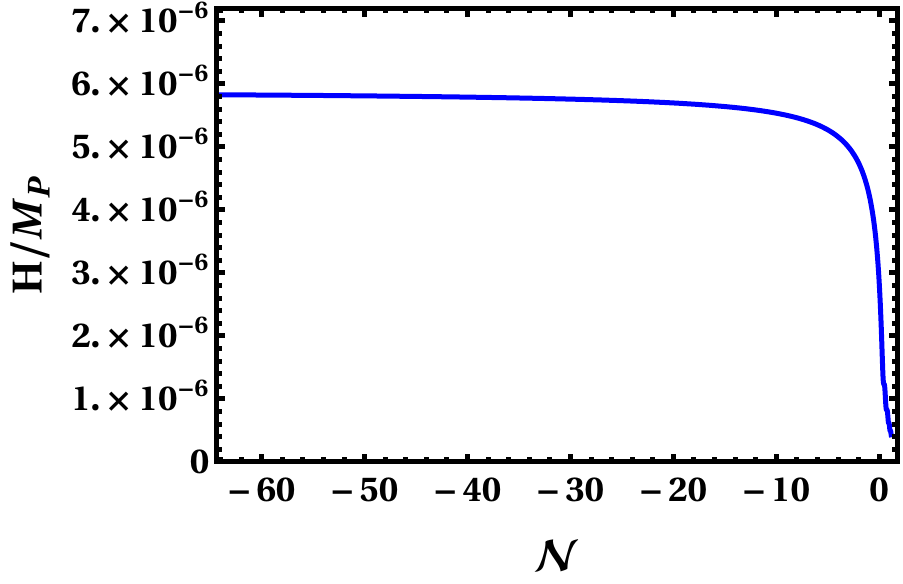}
	\end{subfigure}%
	\begin{subfigure}{.43\textwidth}
		\centering
		\includegraphics[width=.82\linewidth]{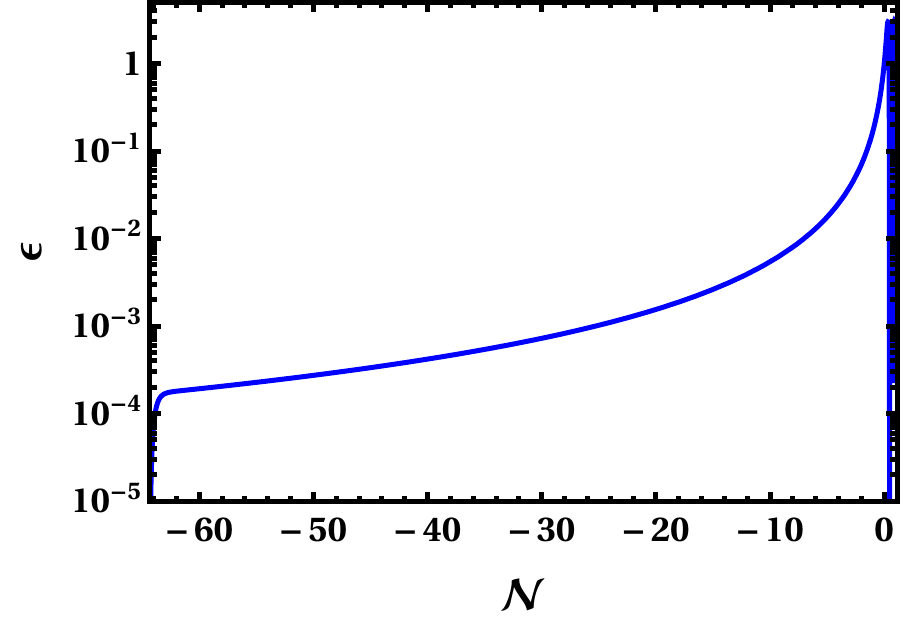}
	\end{subfigure}
	%
	%
	\caption{The evolution of $ H/\mpl$ (left) and $\epsilon$ (right) with respect to $\mathcal{N}$ for BP$a$.}
	\label{hubbleepsi}
\end{figure*}

The background field evolutions are obtained by solving the
Eqs.~\eqref{fildeq} and \eqref{hubble1} with the initial field values at $ t=t_{\text{in}} $ 
providing $e$-folding number between CMB pivot scale and the time at the end of  inflation
$ \Delta \mathcal{N}_{\text{CMB}} \equiv \ln \frac{a(t_{\text{end}})}{a(t_{\text{CMB}})} $ larger than about 50-60. 
In what follows we set $\mpl=1$.

We show the time evolution of the background field $\bar{\phi}^1 = \varphi$ for BP$a$ in the left
panel of Fig.~\ref{fieldevoR} in blue solid lines. In the right panel of Fig.~\ref{fieldevoR} we plot the evolution of the background fields
of $\rho_1$, $\rho_2$ and $\rho_3$ by dot-dashed, dotted and dashed lines respectively.
For the sake of illustration here we only provide
figures for BP$a$ however we have checked other BPs also produce similar trajectories and inflationary dynamics. 
The evolutions of $H$ (in $\mpl$ unit) and
$\epsilon$ are displayed in Fig.~\ref{hubbleepsi} in the left and right panels
respectively. Inflation ends via breakdown of slow-roll condition i.e. 
when $\epsilon(t_{\text{end}}) = 1$.

Instead of $t$, we interchangeably use the number of $e$-foldings before the end of inflation
\begin{align}
	 \mathcal{N} \equiv \ln \frac{a(t)}{a(\tnd)}
\end{align}
as a cosmological evolution variable to understand the inflationary dynamics. With this definition, $\tnd$ corresponds 
to zero $e$-foldings, whereas negative and positive $\mathcal{N}$
denote the amount of $e$-foldings before and after the end of inflation respectively. 

At the pivot scale $ k = k_{*} $, the amplitude of $\mathcal{P}_{\mathcal{R}}(k)$ should match the scalar amplitude 
measurement of Planck 2018 $A_{s} = (2.099\pm 0.014)\times 10^{-9}$ at 68\% CL~\cite{Planck:2018jri}.
We find that the pivot scale $k_*$ exit horizon 
at around $\mathcal N \sim - 57$ for all BPs. 
However, it should be reminded that the relation between the number of $e$-foldings before the end of inflation and the pivot scale $k_*$
depends on the thermal history after inflation.

\begin{figure}[htbp]
	\center
	\includegraphics[width=.4 \textwidth]{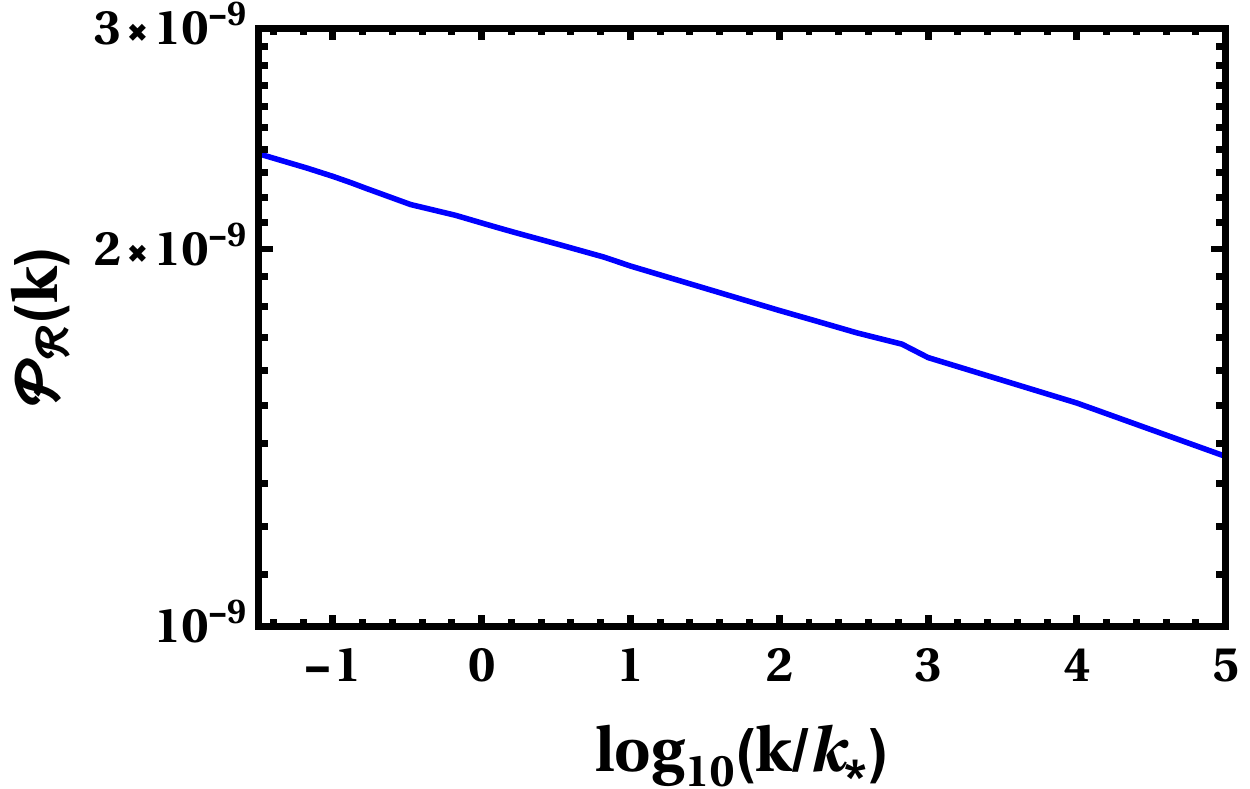}
	\caption{The power spectrum  $\mathcal{P}_{\mathcal{R}}(k)$ for the curvature 
	perturbation as given in Eq.~\eqref{eq:PR} for BP$a$ for illustration.} 
	\label{PRplot}
\end{figure}

\begin{figure}
\centering
	\includegraphics[width=.4 \textwidth]{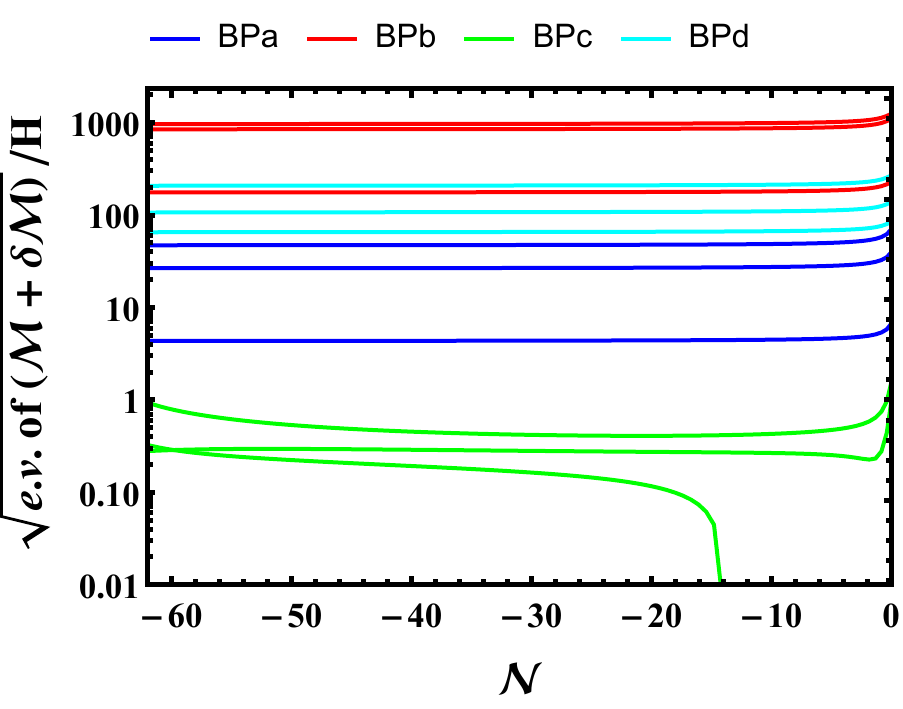}
	\caption{The square roots of the eigenvalues of $ \mathcal{M}^{I}_{J} + \delta \mathcal{M}^{I}_{J} $ 
	in Eq.~\eqref{eq:fluc} for three heavy modes for four BPs during the inflation normalized by the Hubble parameter $ H $.}
	\label{Fig:massspectrum}
\end{figure}

\begin{figure*}[htbp]
	\center
	\includegraphics[width=.38 \textwidth]{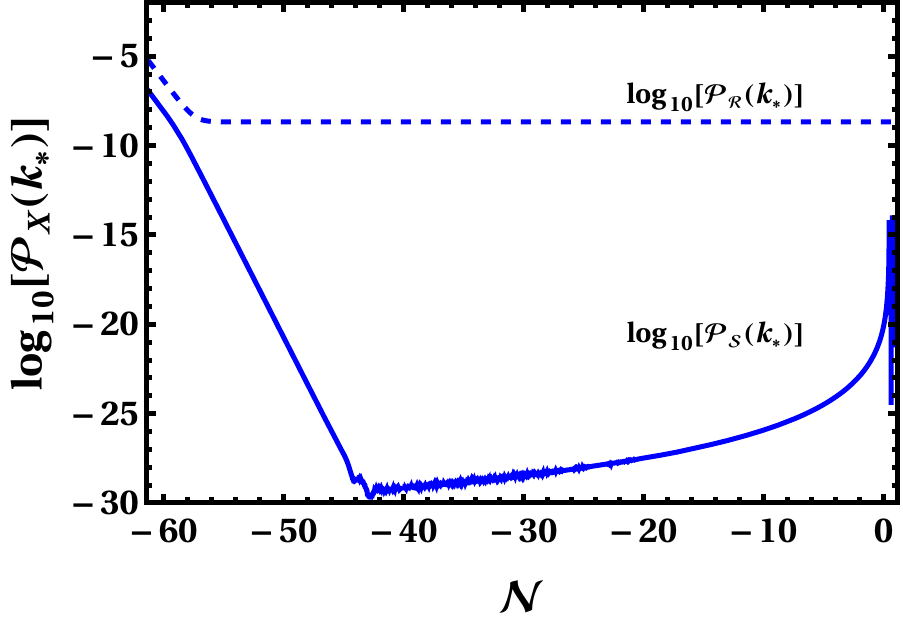}
	\includegraphics[width=.38 \textwidth]{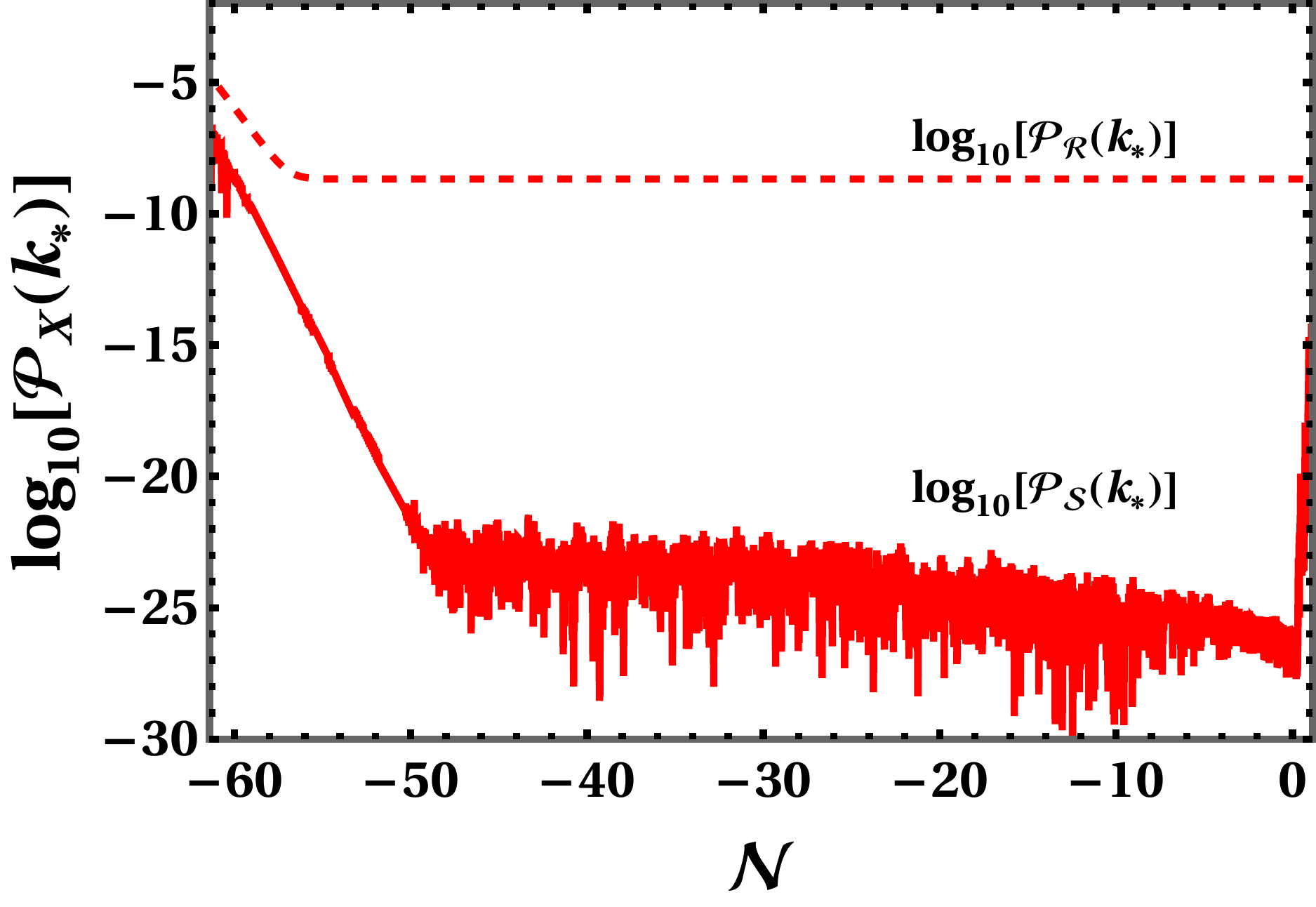}
	\includegraphics[width=.38 \textwidth]{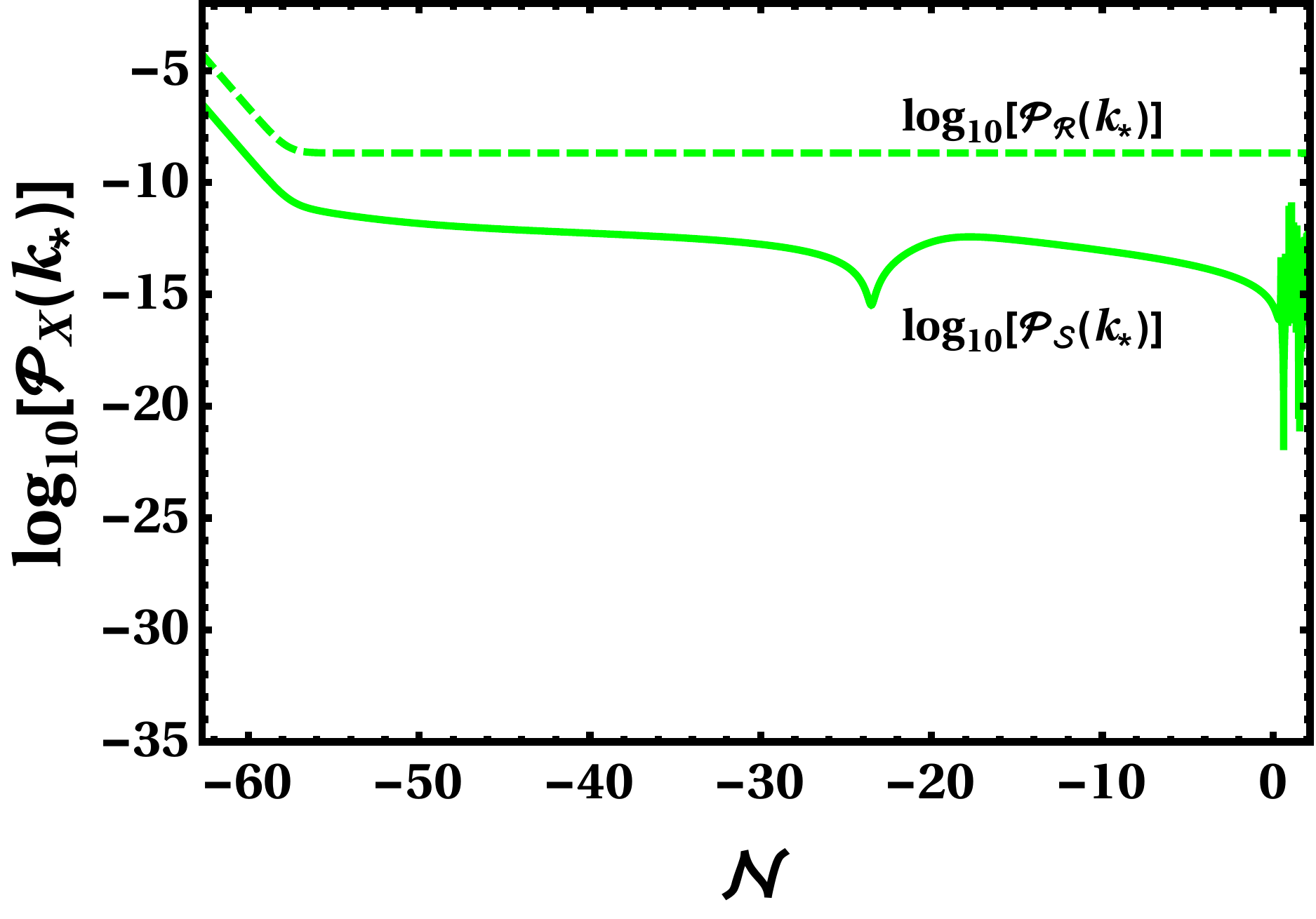}
	\includegraphics[width=.38 \textwidth]{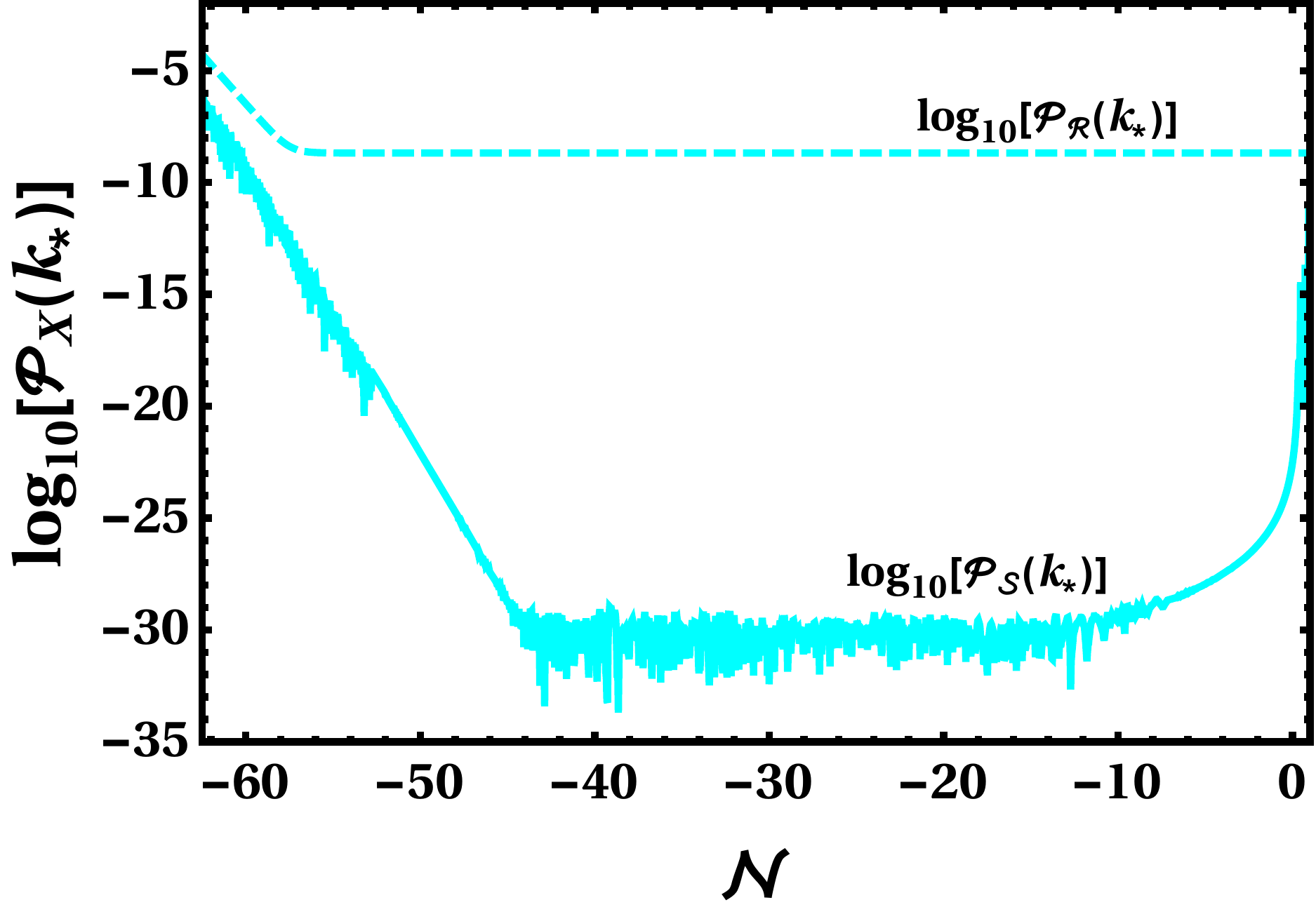}
	\caption{Evolution of the power spectrum of adiabatic mode $\mathcal{P}_{\mathcal{R}}(k)$ and 
		entropy mode $\mathcal{P}_{\mathcal{S}}(k)$ for pivot scale $k_*$ for BP$a$ (upper left), BP$b$ (upper right), BP$c$ (lower left) and
		BP$d$ (lower right) respectively.} 
	\label{PRPSplot}
\end{figure*}

In Fig.~\ref{PRplot} we plot the power spectrum of the curvature perturbation $\mathcal{P}_{\mathcal{R}}(k)$ vs $\log_{10}(k/k_*)$
for BP$a$, which shows nearly scale invariant but clearly red-tilted nature.
We find that the entropy
perturbations $\mathcal{P}_{\mathcal{S}}(k)$, $\mathcal{P}_{\mathcal{U}}(k)$ and $\mathcal{P}_{\mathcal{V}}(k)$ to be 
tiny during inflation blue for BP$ a $, BP$ b $, and BP$ d $.

Indeed, this can be seen from the fact that  the square roots of the eigenvalues of the mass matrix 
$ \mathcal{M}^{I}_{J} + \delta \mathcal{M}^{I}_{J} $ in Eq.~\eqref{eq:fluc} for three modes are 
heavier than the Hubble scale during the inflation for each BPs, as depicted in Fig.~\ref{Fig:massspectrum}. 
These correspond to entropy modes and this implies that fluctuations of the entropy modes are exponentially suppressed during the inflation. 
Also, in these kind of parameters,  the valley approximations can be adopted in which,  by integrating out heavy modes, 
and inflation dynamics can be described by a single field-like one. Some details of the valley approximation is given in Appendix~\ref{Appendix:Valley}.

On the other hand, for BP$ c $, one can see that the masses of other modes 
other than adiabatic one are almost the same or smaller than the Hubble scale. For parameters with
light masses like BP$ c $, one generally cannot adopt the valley approximations, and one in principle has to solve all background and perturbation equations exactly. However, 
we explicitly checked that the isocurvature power spectra for BP$c$ are not exponentially 
suppressed during inflation, and still does not affect the adiabatic fluctuation significantly.
Therefore even with the parameter set such as BP$c$, we can calculate the inflationary observables in the same manner as the single-field case.

For explicit comparison we also plotted the evolution of the power spectra for the adiabatic mode $\mathcal{P}_{\mathcal{R}}(k)$ and the entropy mode 
$\mathcal{P}_{\mathcal{S}}(k)$  in Fig.~\ref{PRPSplot} for all four BPs. The figure illustrates that for all 
BPs the power spectrum $\mathcal{P}_{\mathcal{R}}(k)$ remains much larger 
that of $\mathcal{P}_{\mathcal{S}}(k)$. We have checked this is also true for the $\mathcal{P}_{\mathcal{U}}(k)$ and 
$\mathcal{P}_{\mathcal{V}}(k)$. This should be compared with the corresponding eigenvalues of the mass matrix for each BPs
in Fig.~\ref{Fig:massspectrum}.As mentioned above, for PB$c$, the mass eigenvalues for isocurvature modes 
are not heavier than the Hubble scales, which explains the behavior that the size of ${\cal P}_{\cal S} (k) $ is 
relatively large, although still smaller that the adiabatic one,  
compared to the counterpart in other BPs. However, we emphasize that even in the case of BP$c$ the effects 
of isocurvature modes on the adiabatic one are small enough such that the single-field description is valid. 
We remark that the amplification of the entropy modes such as $\mathcal{P}_{\mathcal{S}}(k)$ 
at around the end of inflation can happen as can be  seen in Fig.~\ref{PRPSplot}, which  might have originated from 
preheating after inflation (see e.g. Refs.~\cite{Bassett:1999ta,Liddle:1999hq,Gordon:2000hv}). 
We leave out a detailed analysis on this issue  for future work.

\begin{figure}[htbp!]
	\centering
	\includegraphics[width=.4\textwidth]{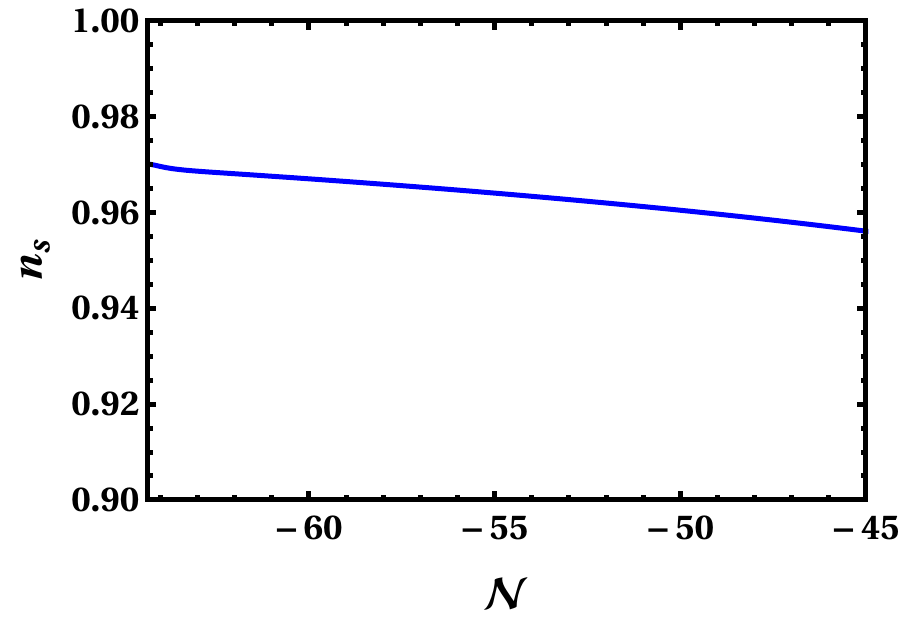}
	\caption{The $n_{\mathcal{R}}$ vs $\mathcal{N}$ plot as in Eq.~\eqref{specin2} for BP$a$.} 
	\label{nRplot}
\end{figure}

Finally, we plot the spectral index $n_{s}$
in Figure~\ref{nRplot}. As the entropy perturbations are tiny, while finding Fig.~\ref{nRplot}, we simply utilize the
approximate expression given for the single field inflation in Eq.~\eqref{specin2}. We find that for $\mathcal{N}=-57.5$ and $-57.2$
(i.e. at $t=t_*$) the spectral indices for all the BPs match with the Planck 
2018 observation i.e. $n_{s}= 0.9649\pm 0.0042$ at 68\% CL~\cite{Planck:2018jri} as also can be seen from Fig.~\ref{nRplot}. 
The Planck 2018 data also obtained the bound for the tensor-to-scalar ratio as $r< 0.056$~\cite{Planck:2018jri}. 
By including the BICEP/Keck 2018 data, the constraint became tighter as $r< 0.036$~\cite{BICEP:2021xfz}.

We find $r\approx 3.35\times 10^{-3}$
and $\approx 3.37\times 10^{-3}$ for the respective BPs, which is well below 
the current observational bounds, but can be detectable future CMB B-mode experiments 
such as LiteBIRD \cite{Matsumura:2013aja} and the Simons Observatory \cite{SimonsObservatory:2018koc}.
Although we do not discuss in detail and provide any figures for other PBs, we have checked that the other cases almost give similar values for $n_{s}$ and $r$.

\section{Implications for collider experiments}\label{coll}
\begin{figure*}[htbp]
\centering
\includegraphics[width=.44 \textwidth]{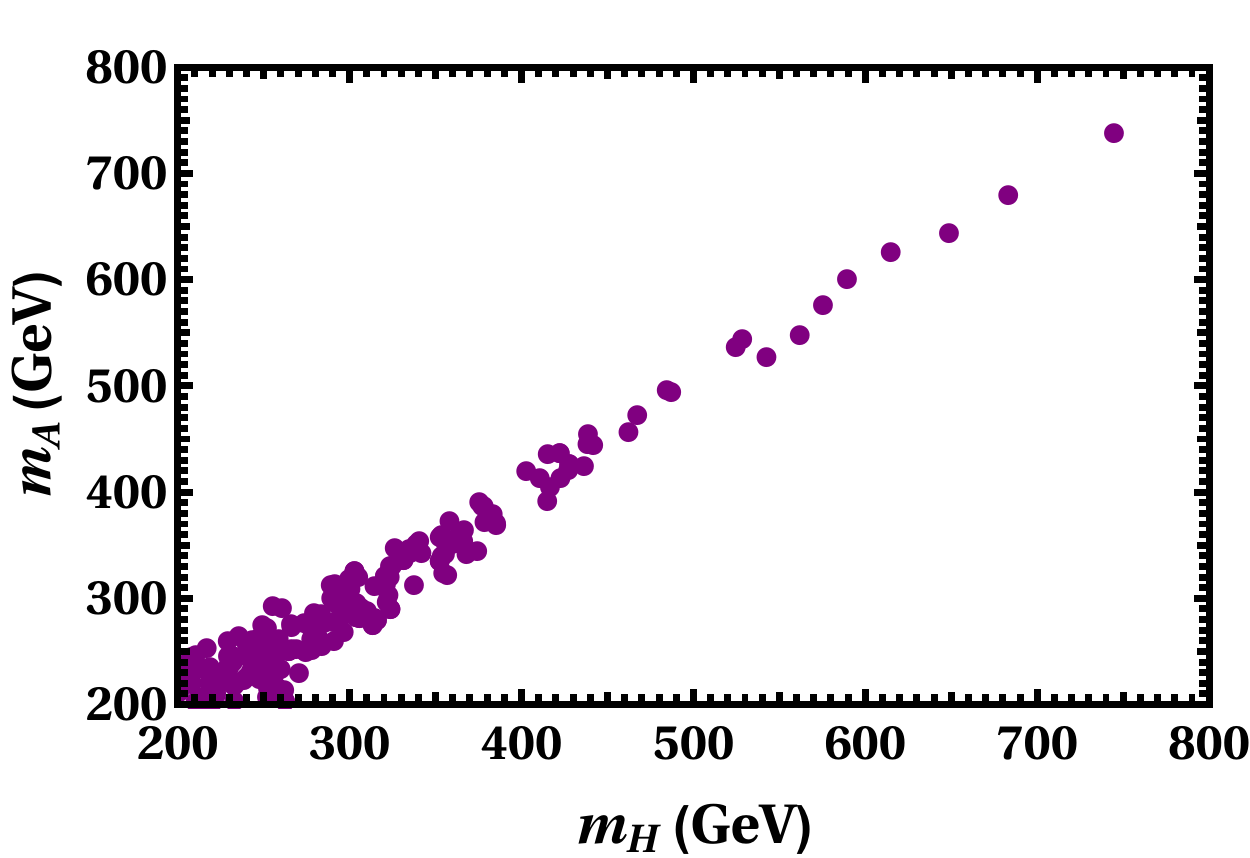}
\includegraphics[width=.44 \textwidth]{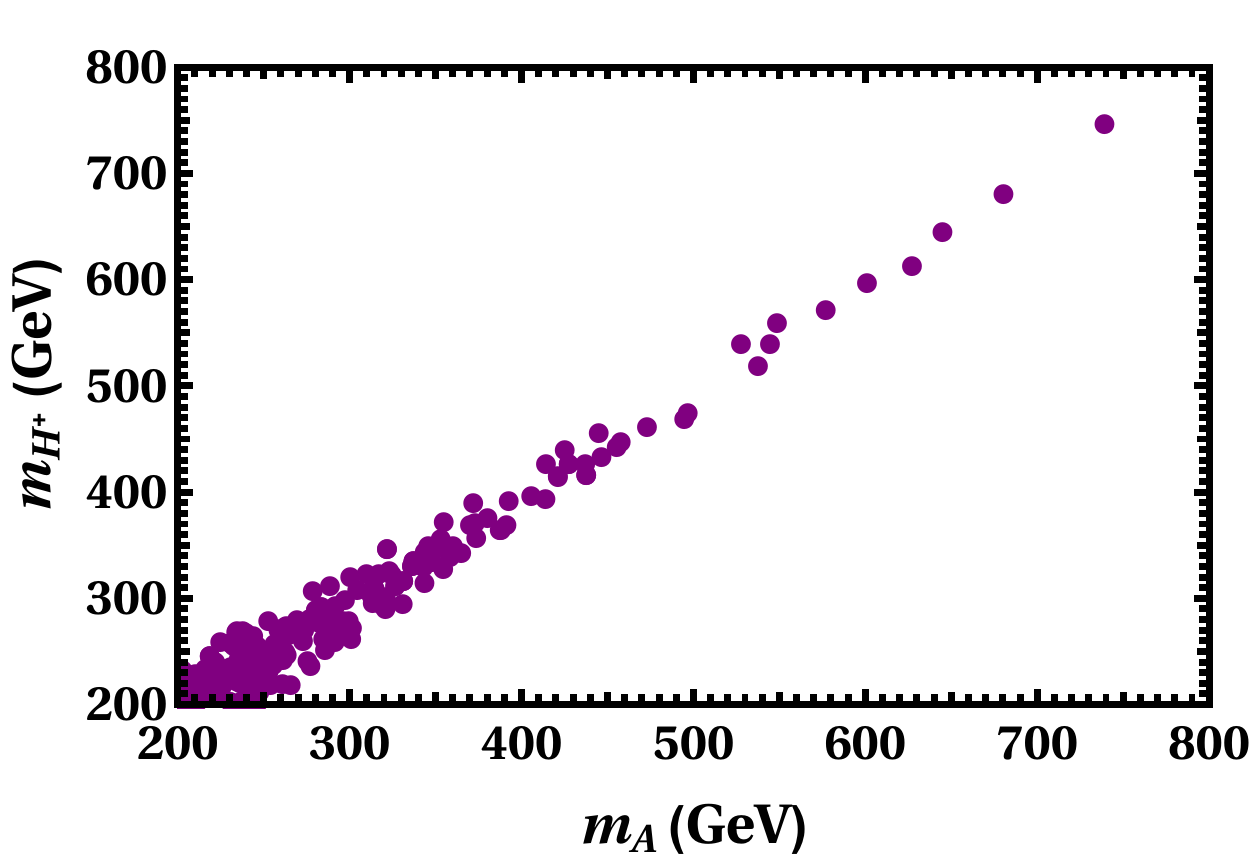}
\caption{The scanned parameter space in the $m_A$--$m_H$ (left) and $m_A$--$m_{H^\pm}$ (right)
plane that can provide successful $R^2$-Higgs inflation.}
\label{scannedmass}
\end{figure*}

Let us discuss implications of the $R^2$-Higgs inflation for collider experiments. For illustration, in Sec.~\ref{resul}, we have chosen 
benchmark points for our analysis. Notwithstanding, there exists larger sub-TeV parameter space for $\mathsf{H}$, $A$ and $\mathsf{H}^\pm$
that can account for $R^2$-Higgs inflation in the g2HDM. In Fig.~\ref{scannedmass} we provide scanned parameter space for 
$m_A$, $m_H$ and $m_{H^\pm}$ that can provide successful $R^2$-Higgs inflation satisfying all inflationary
conditions and observational constraints from Planck 2018~\cite{Planck:2018jri}.

As can be seen from Fig.~\ref{scannedmass} for the successful $R^2$-Higgs inflation quasi-degenerate mass spectrum is required for the heavy 
Higgs bosons $m_\mathsf{H}$, $m_A$ and $m_{\mathsf{H}^\pm}$. This finding is similar
to Higgs inflation in g2HDM but without the $R^2$ term~\cite{Modak:2020fij}.
This is primarily due to the requirement of perturbativity for the $\tilde \eta_i$ 
for inflationary dynamics at high scale. 
We find that parameter points with $\eta_i > 1$ at low scale (i.e. $y=0$) 
grow too large at high scale and get excluded by the perturbativity requirements.
Due to limited computational facility for scanning we restricted all $\eta_i$ at low scale to be  $<1$.
With a common $\mu_{22}^2$ terms, the $m_\mathsf{H}$, $m_A$ and $m_{\mathsf{H}^\pm}$ mass degeneracy gets practically 
restricted due to these small values of $\eta_i$ at low scale which can be seen easily from Eqs.~\eqref{eq:mHh}, \eqref{eq:mA} and \eqref{eq:mHpm}.
This has unique implications for collider experiments, that is,
a future discovery of quasi-degenerate $m_\mathsf{H}$, $m_A$ and $m_{\mathsf{H}^\pm}$ would provide a
smoking gun signature for $R^2$-Higgs inflation in the g2HDM.

While the Yukawa couplings $\rho^F_{ij}$ do not play significant role in the inflationary dynamics and
only enter in the $\beta$ functions of the quartic couplings $\eta_i$s, however, they could play important role 
in the discovery and/or constraining the parameter space for $\mathsf{H}$, $A$ and $\mathsf{H}^\pm$. Here
we assumed $\rho^F_{ii}\sim \lambda^F_i$ with suppressed off diagonal elements for the $\rho^F_{ij}$ matrices.
In particular, we assumed extra Yukawa couplings $\rho^U_{tt}=0.5$ and $\rho^U_{tc}=0.2$ for  the RG running 
for all the BPs discussed in the previous section and turned off other extra Yukawa couplings for simplicity.
In what follows we shall see that for these values of extra Yukawa couplings are allowed by direct and indirect searches 
and may lead to discovery of the heavy Higgs bosons. 

\subsection{Indirect searches}
First we focus on the coupling measurements $h$ boson at the LHC. A nonvanishing 
$c_\gamma$ can alter the couplings of 125 GeV $h$ boson e.g. to fermions, as can be seen from Eq.~\eqref{eff}.
Following the prescription given in Ref.~\cite{Hou:2018uvr} 
we find that $|\rho^U_{tt}|=0.5$ is well allowed at $2 \sigma$ by the current measurements of top Yukawa coupling
of $h$ by ATLAS~\cite{ATLAS:2020qdt} and CMS~\cite{CMS:2020gsy} with full Run 2 data. The limit is rather
weak primarily due to the small $c_\gamma$ values (see Table~\ref{lowscale}) for all the BPs.
We remark that such coupling measurements in general allows $\rho^U_{tt}\sim 0.5$ if $c_\gamma \lesssim 0.1$.

The $\rho^U_{tt}$ also receives stringent constraints from flavor physics,  e.g. 
nonvanishing $\rho^U_{tt}$ enters in $B_{s,d}$ mixing amplitude as well as branching 
ratio of $B\to X_s\gamma$ ($\mathcal{B}(B\to X_s\gamma)$) at one loop through 
$tb\mathsf{H}^\pm$ vertex~\cite{Altunkaynak:2015twa}. The strongest limit arises however from the $B_{s,d}$ mixing.
Allowing $2\sigma$ error on the UTfit results for $B_{s,d}$~\cite{UTfitBsmix} and following the expression
given in Ref.~\cite{Altunkaynak:2015twa}, we find that $|\rho^U_{tt}|\lesssim0.5$ is allowed at $2\sigma$ for all the BPs. 
This suggest that for the ballpark value of $\rho^U_{tt}$ assumed here, flavor physics already provides
indirect probe for the inflationary dynamics in particular for $m_{\mathsf{H}^\pm}\sim 200$--600 GeV, but the constraint becomes milder
for heavier $m_{\mathsf{H}^\pm}$. In this regard future LHCb~\cite{LHCb:2018roe} and Belle-II~~\cite{Belle-II:2018jsg}
measurements would offer a further stringent test for the sub-TeV $m_{\mathsf{H}^\pm}$ if $\rho^U_{tt}$ is not vanishingly small.

The flavor changing coupling $\rho^U_{tc}$ does not enter $h$ boson couplings at tree level
however it may induce top flavor changing decay $t\to c h$ if $c_\gamma$ is nonzero. 
Such searches are performed and strong upper limits on the branching ratios of $t\to c h$ ($\mathcal{B}(t\to c h)$) are already set 
by both ATLAS~\cite{Aaboud:2018oqm} and CMS~\cite{CMS:2021bdg}. We find
that the CMS 95\% CL upper limit $\mathcal{B}(t\to c h)< 7.3\times 10^{-4}$~\cite{CMS:2021bdg}
is mildly stronger than the ATLAS one. Utilizing these limits it has been 
found that $\rho^U_{tc} \lesssim 0.75$ is still allowed at 95\% CL if $c_\gamma=0.1$~\cite{Hou:2020tnc}.
This means that our chosen value $\rho^U_{tc}=0.2$ is well allowed by data. There also exist constraints on $\rho^U_{tc}$ from 
flavor physics. Relevant constraints arise also from $\mathcal{B}(B\to X_s\gamma)$ where $\rho^U_{tc}$ enters via charm loop through
$\mathsf{H}^+$ coupling~\cite{Altunkaynak:2015twa}. 
Reinterpreting results from Ref.~\cite{Crivellin:2013wna} we find that $\rho^U_{tc} \gtrsim 1$  is excluded at $2\sigma$
if $m_{\mathsf{H}^\pm}\sim 200$--500 GeV. We remark that the constraint is weak and becomes even milder for heavier  $m_{\mathsf{H}^\pm}$.

In general other $\rho^F_{ij}$  couplings such as $\rho^D_{bb}$ and $\rho^U_{tu}$ could still be large, 
e.g., extra Yukawa couplings $|\rho^D_{bb}|\lesssim 0.1$--$0.15$ is still allowed by  current 
data for $m_\mathsf{H},m_A,m_{\mathsf{H}^\pm} \in[200,800]$ GeV~\cite{Modak:2018csw,Modak:2019nzl,Modak:2020uyq}.
Furthermore, we also remark that there also exist some indirect measurements that provide some constraints on $\rho^U_{tu}$. E.g., 
$B\to \mu \nu$  and $D$-meson mixing provide some constraints but still allow $\rho^U_{tu}\sim 0.1-0.2$ at $2\sigma$ 
level~\cite{Hou:2019uxa,Hou:2020ciy}. If they are nonvanishing they may offer additional probes for the parameter space
required for $R^2$-Higgs inflation in the g2HDM.

\subsection{Direct searches}
Nonzero $\rho^U_{tt}$ can induce $V_{tb}$ enhanced $bg\to \bar t \mathsf{H}^+$ and 
$gg\to \bar t b \mathsf{H}^+$ processes (charge conjugate processes are implied). 
The processes $pp\to \bar t (b) \mathsf{H}^+$ followed by $\mathsf{H}^+\to t \bar b$ are the 
conventional search program for the $\mathsf{H}^\pm$ of ATLAS~\cite{ATLAS:2020jqj} and CMS~\cite{Sirunyan:2020hwv}. 
Further for $m_A/m_\mathsf{H} > 2 m_t$, $\rho^U_{tt}$ coupling can initiate $gg\to \mathsf{H}/A \to t \bar t$, which are already being searched by
ATLAS~\cite{ATLAS:2017snw} and CMS~\cite{CMS:2019pzc}. In general, such searches exclude $\rho^U_{tt}\gtrsim0.6$--1 at 95\% CL for 
$m_\mathsf{H},m_A,m_{\mathsf{H}^\pm} \in[200,800]$ GeV~\cite{Ghosh:2019exx}.

There also exist direct searches that can constrain the flavor changing coupling $\rho^U_{tc}$. The most relevant search in this 
regard is CMS search for SM four-top production~\cite{CMS:2019rvj}. It has been found~\cite{Hou:2018zmg,Hou:2019qqi} 
that $\rho^U_{tc}$ coupling induced $cg\to t \mathsf{H}/tA \to t t \bar c$ processes contribute abundantly to the control
region of $t\bar t W$ background of the CMS search which excludes $|\rho^U_{tc}|\lesssim 0.4$--0.6 in the 
$m_\mathsf{H},m_A\in[200,600]$ GeV~\cite{Kohda:2017fkn,Hou:2018zmg,Hou:2020tnc,Hou:2019qqi,Hou:2019mve,Hou:2019gpn,Ghosh:2019exx,Hou:2020chc,Hou:2021xiq}. 
As our working assumption was $\rho^F_{ii}\sim \lambda^F_i$ and suppressed off-diagonal elements, in general couplings 
such as $\rho^D_{bb}$ and $\rho^L_{\tau\tau}$ are below the sensitivity of the LHC.

\subsection{Probing the quasi-degeneracy} 

The processes mentioned above together may allow discovery of the heavy Higgs
bosons $\mathsf{H}$, $\mathsf{H}^\pm$ and $A$, however, one could only attribute a parameter space in the g2HDM
to the $R^2$-Higgs inflation if quasi-degeneracy is also observed.
This would require tricky reconstruction of the masses of these heavy Higgs
bosons or finding out processes that are sensitive to mass degeneracies. In this subsection we discuss
how to probe such quasi-degeneracy in LHC or future lepton colliders. 

For nonvanishing $\rho_{tt}$ the $\mathsf{H}^\pm$ can be reconstructed in the sub-TeV range
via $bg\to \bar t (b) \mathsf{H}^+$ followed by $\mathsf{H}^+\to t \bar b$ decay
as already discussed by ATLAS~\cite{ATLAS:2020jqj} and CMS~\cite{Sirunyan:2020hwv}.
In general reconstruction might be also possible e.g. via 
process such as $gg\to \mathsf{H}/A \to t \bar t$ if $m_A/m_\mathsf{H} > 2 m_t$. 
The searches performed so far by ATLAS~\cite{ATLAS:2017snw} and CMS~\cite{CMS:2019pzc} 
assume decoupled $m_A$ and $m_\mathsf{H}$. Therefore while discovery is possible, however, extraction of information on quasi-degeneracy
would be particularly difficult due to interference between $gg\to A \to t \bar t$,
$gg\to \mathsf{H} \to t \bar t$ and SM $gg\to t \bar t$. For nonvanishing $\rho^U_{tc}$
and $\rho^U_{tt}$ one may have discovery via $gg \to \mathsf{H}/A \to t \bar c $~\cite{Altunkaynak:2015twa},
however the interference between $gg \to A \to t \bar c $ and $gg \to \mathsf{H} \to t \bar c $
would again obscure the information on mass degeneracy. Additionally one may have discovery 
via $cg\to t \mathsf{H}/tA \to t t \bar t$~\cite{Kohda:2017fkn} or 
$cg \to b H^+ \to b t  \bar b$~\cite{Ghosh:2019exx} at the high-luminosity LHC if both 
$\rho^U_{tc}$ and $\rho^U_{tt}$ are nonzero.

It is clear that to probe quasi-degeneracy of $\mathsf{H}$ and $A$ one
requires careful analysis due to multiple interfering contributions. In such scenarios we propose to study
$cg\to t \mathsf{H}/tA \to t t \bar c$ (denoted as same-sign top) at the LHC which may provide 
\textit{smoking gun} signature for the quasi-degeneracy between $\mathsf{H}$ and $A$.
It has been found that if $\mathsf{H}$ and $A$ are both mass and width degenerate the process 
$cg\to t \mathsf{H} \to t t \bar c$  and $cg\to tA \to t t \bar c$ cancel each other exactly due to destructive
interference~\cite{Kohda:2017fkn}. This is primarily due to the amplitude for $cg\to tA \to t t \bar c$
picks up a factor of $i^2 \gamma_5$ compared to $cg\to t \mathsf{H} \to t t \bar c$, as can be seen from Eq.~\eqref{eff}.
The cancellation diminishes if the mass and/or widths become non-degenerate.


Let us briefly discuss the potential of the same-sign top signature to probe
quasi-degeneracy between $\mathsf{H}$ and $A$.
For illustration we consider BP$a$ and BP$c$.
Moreover, we assume $\rho^U_{tt}=0.5$ and $\rho^U_{tc}=0.5$ which we have checked are allowed by all
 direct and indirect searches mentioned above. We turn off all other $\rho_{ij}$ couplings, however shall return 
to their impact on mass reconstruction at the end of this section. Under the above mentioned assumptions
the total decay widths for $A$ ($\mathsf{H}$) are sum of partial rates of $A\to t \bar c + \bar t c$
($\mathsf{H}\to t \bar c + \bar t c$) and, $A\to \bar t t$  ($\mathsf{H}\to \bar t t$) for  BP$a$.
But for BP$c$ both $A$ and $\mathsf{H}$ decays practically $100\%$ to $t \bar c + \bar t c$.
For $\rho^U_{tt}=0.5$ and $\rho^U_{tc}=0.5$ we find the decay widths of $A$ and $\mathsf{H}$  are 2.43 (8.58) and 2.41 (6.04) GeV for 
BP$c$ (BP$a$).

The same-sign top can be searched at LHC via $pp \to tH/tA + X \to tt\bar c + X$ with both the top
quarks decaying semileptonically comprising same-sign dilepton ($ee$, $e\mu$, $\mu\mu$) plus at least three jets
with at least two $b$-tagged and one non-$b$-tagged, and missing energy ($E_T^{\rm miss}$)

The SM backgrounds for the process are $t\bar t Z$, $t\bar t W$, $4t$, $t\bar t h$ and $tZ +$ jets.
Additionally, for the same-sign top signature the SM $t\bar t$ and $Z/\gamma^* +$ jets processes
would contribute if one of the lepton charge is misidentified ($Q$-flip). Notwithstanding,
it has been found that the non-prompt background could be $\sim 1.5$ times of the 
$t\bar t W$ background for the same-sign top signature~\cite{Kohda:2017fkn}.

In order to demonstrate the discovery potential we generate the signal and background events at $\sqrt{s}=14$ TeV via
MadGraph5\_aMC@NLO~\cite{Alwall:2014hca} with the parton distribution function (PDF) set NN23LO1~\cite{Ball:2013hta} .
The events are then interfaced with 
PYTHIA~6.4~\cite{Sjostrand:2006za} for showering and hadronization, and 
then fed into Delphes~3.4.2~\cite{deFavereau:2013fsa} to incorporate detector effects (ATLAS based).

To suppress backgrounds and optimize for the same-sign top 
signature we apply following event selection cuts. The leading and subleading 
lepton transverse momenta $p_T$ should be  $>25$  and $> 20$ GeV respectively,
while the pseudo-rapidity $|\eta| < 2.5$. 
For all three jets we require $p_T> 20$\;GeV and also $|\eta| < 2.5$, 
and $E^{\rm miss}_{T} > 30$\;GeV.
The separation $\Delta R$ between any jets and a lepton ($\Delta R_{\ell j}$), 
the two $b$-jets ($\Delta R_{bb}$), 
and any two leptons ($\Delta R_{\ell\ell}$)
should be $\Delta R > 0.4$. 
Finally, we impose $H_T$ i.e. the sum of the  $p_T$ of the two leading leptons included and two leading $b$-jets and 
the leading non $b$-tagged jets should $ > 300$\;GeV. 

The background cross sections after the application of 
the above selection cuts are summarized in Table~\ref{sstsigbackg} while the signal cross sections for the reference mass scenario 
BP$a$ (BP$c$) is 0.023 (0.18) fb.  The corresponding statistical significances are 
$\sim 1 \sigma$ and $\sim 4 \sigma$ respectively with 3000 fb$^{-1}$ luminosity; which are
estimated by using $\mathcal{Z} = \sqrt{2[ (S+B)\ln( 1+S/B )-S ]}$~\cite{Cowan:2010js},
where the $S$ and $B$ are the number of signal and background events after selection cuts.
This simply illustrates that discovery of same-sign top process is not possible for both the scenarios even at the high luminosity LHC (HL-LHC).
In general, same-sign top signature for these reference mass ranges are expected to be discovered
much earlier than full HL-LHC data for $\rho^U_{tt}=0.5$ and $\rho^U_{tc}=0.5$ if $\mathsf{H}$ and $A$ are degenerate~\cite{Kohda:2017fkn}. 
Hence, discoveries of $gg\to \mathsf{H}/A \to t \bar t$, $cg\to t \mathsf{H}/tA \to t t \bar t$ and $cg\to b H^+ \to b t \bar b$  
and non-observation or milder significance of the same-sign top in the HL-LHC era may indicate quasi-degeneracy of
$\mathsf{H}$ and $A$ whereas, the charged Higgs mass can be reconstructed 
via $bg\to \bar t (b) \mathsf{H}^+ \to \bar t (b) t \bar b$.

\begin{table}[htbp!]
\centering
\begin{tabular}{c |l c c c }
\hline

                     \,Backgrounds\,                 & \ \,Cross section (fb)\,     
& \\
\hline
\hline
                       $t\bar{t}W$                  &  1.31               \\
                       $t\bar{t}Z$                  &  1.97               \\
                       $4t$                         &  0.316               \\
                       $tZ+$\,jets                  &   0.255                \\
                      $t\bar t h$                   &   0.07                \\                       
                     $Q$-flip                       &   0.024                 \\
                     nonprompt                      & $1.5\times t\bar{t}W$      \\
\hline
\hline
\end{tabular}
\caption{The background cross sections after selection cuts for the same-sign top search.
The backgrounds cross sections are all normalized to either NLO or NNLO as in Ref.~\cite{Kohda:2017fkn}.
}
\label{sstsigbackg}
\end{table}


%

Probing the quasi-degeneracy at the LHC becomes particularly challenging if $\rho^U_{tc}$ or $\rho^U_{tt}$ are small.
Furthermore, processes such as $gg\to \mathsf{H}/A \to t \bar t$ and $cg\to t \mathsf{H}/tA \to t t \bar t$ 
are only sensitive above $m_A/m_\mathsf{H} > 2 m_t$ threshold. In such cases $e^+e^-$ colliders such as ILC or FCC-$ee$ 
could be useful for discovery and, possibly even for probing quasi-degeneracy.
In this regard we propose to study $e^+e^-\to Z^* \to A \mathsf{H}$, $e^+e^-\to Z^*/\gamma \to \mathsf{H}^+ \mathsf{H}^-$, 
$e^+e^-\to Z^* \to A h$ followed by $A/\mathsf{H} \to t \bar c + \bar t c/ t \bar t$ 
or $\mathsf{H}^+ \to c \bar b / t \bar b$. Depending on the values of $\rho^U_{tc}$ or $\rho^U_{tt}$, these processes may 
require $\geq 1$ TeV CM energy and/or high-luminosity $e^+e^-$ collider for discovery, while
probing the the quasi-degeneracy of $\mathsf{H}$ and $A$, $\mathsf{H}^\pm$ would perhaps require 
even higher statistics.

So far we have turned off other $\rho_{ij}$ couplings for simplicity.
In general $\rho_{bb}^D$ could be nonvanishing and would open up new modes for mass reconstruction 
such as $bg \to b A \to b Z h$ process at the LHC or at future lepton collider via $e^+e^-\to Z^* \to A \mathsf{H}$,
followed by $A (\mathsf{H}) \to b \bar b$ decay. For nonzero 
$\rho^L_{\tau \tau}$ discovery is possible via $gg \to \mathsf{H}/A\to  \tau^+ \tau^-$ at the LHC 
or $gb \to t H^+ \to t \tau^+ \nu_\tau$ processes. For finite discussion, we however do not turn on 
all these $\rho_{ij}$ couplings together since they would initiate many new direct and indirect signatures that are
not discussed here. Such scenario would nonetheless be interesting and require a more dedicated analysis 
which is beyond the scope of the current paper.

\section{Discussion and Summary}\label{disc}

We have studied $R^2$-Higgs inflation in the g2HDM  
where the inflationary dynamics consists of four fields $\varphi$, $\rho_{1}$, $\rho_{2}$ and $\rho_{3}$ using the covariant formalism.
We first discussed relevant background dynamics and perturbation theory for the field fluctuations for
our four field model. We found that, by numerically solving the set of equations
for the background and perturbation evolutions,  primordial power
spectra  for the parameter sets consistent with Planck observations~\cite{Planck:2018jri}  and low energy constraints can be well 
described by a single-field approximation where the field $\varphi$ nearly plays the role 
of inflaton, whereas $\rho_1$, $\rho_2$ and $\rho_3$ play isocurvature fields during inflation and those isocurvature
modes scarcely affect the adiabatic one by appropriately 
choosing the initial values for isocurvature fields.
However, we note that there may exist parameter space 
where the entropy modes affect the power spectrum for 
the adiabatic one and/or primordial non-Gaussianities. This shall be studied elsewhere.  

Throughout the paper we have just turned on one nonminimal couplings $\xi_{11}$ for simplicity.
In general the nonminimal couplings $\xi_{12}$ and $\xi_{22}$ can also drive inflation as discussed in Ref.~\cite{Modak:2020fij}.
In the 2HDM inflation without the $R^2$ term, the inflationary dynamics for the nonminimal couplings $\xi_{22}$ (and $\xi_{12}$) is
quite similar to that of $\xi_{11}$~\cite{Modak:2020fij}. However, a similar conclusion can not be drawn here. 
As the parameterization of Eq.~\eqref{filedparam} of the current article is different than the one in Ref.~\cite{Modak:2020fij} 
the different $\xi_{ij}$ couplings may have very distinct inflationary dynamics. While it would indeed be interesting to see the impacts of
these nonminimal couplings individually or, when they are turned on together, however, we leave out a detailed analysis on this for future.

For illustration we chose four  benchmark points for our analysis with $m_\mathsf{H}$, $m_A$ and $m_{\mathsf{H}^\pm} \sim$ 400 GeV.
To satisfy the normalization to CMB power spectrum~\cite{Planck:2018jri}, in the $R^2$-like BP$a$  and $c$ we have 
assumed the scalaron self couplings $\xi_R$ to be large. In the mixed $R^2$-Higgs like BP$b$ and $d$ 
the normalization to CMB data is achieved by considering both $\xi_R$ and nonminimal coupling $\xi_{11}$ to be relatively
large. For all the BPs, the predicted spectral index $n_{s}$ and tensor-to-scalar ratio $r$ are within their experimental bounds~\cite{Planck:2018jri}.

Although for all the benchmark points we considered $m_\mathsf{H}$, $m_A$ and $m_{\mathsf{H}^\pm} \sim$ 400 GeV, 
there exists parameter space for a successful inflationary scenario
in the sub-TeV range i.e. $m_\mathsf{H}$, $m_A$ and $m_{\mathsf{H}^\pm} \in [200,800]$ GeV, as found in Ref.~\cite{Modak:2020fij}.
This mass range has a unique impact for the ongoing collider experiments such as the LHC(b) and Belle-II. 
We discussed a discovery scope for these bosons at the upcoming LHC run and, plausible indirect probes at the flavor 
machines such as LHCb and Belle-II. 
A discovery of these additional bosons along with 
the confirmation of their quasi-degeneracy may hint the g2HDM 
as a likely mechanism for the cosmic inflation. 
Here we also remark that we have assumed all $\rho^F_{ij}$ couplings to be real. In general, 
along with the quartic couplings $\eta_{5,6,7}$ they can be complex in
nature. The implications of such complex couplings during (and after)
inflation including baryogenesis are yet to be analyzed in the g2HDM. (See Ref. \cite{Lee:2020yaj} for
a baryogenesis scenario during the reheating in Higgs inflation.) However, they are already within 
the reach~\cite{Modak:2020uyq} of CP sensitive measurements such as electron electric dipole moment
of ACME collaboration~\cite{ACME:2018yjb} and  the CP asymmetry for $B \to X_s \gamma$ decay at Belle~\cite{Belle:2018iff}. 

We also further remark on the unitarity problem of the 2HDM inflation model. 
The cut-off scale for 2HDM inflation at low field regime is given by
$ \min \left(M_{P}/\xi_{ij}\right) $ with $ i,j=1,2 $~\cite{Gong:2012ri}. 
As already mentioned in the introduction, inflationary dynamics with large 
field values does not suffer the unitarity violation due to 
field-dependent cut-off. However, it is known that the issue of unitarity arises 
again during the preheating stages since the produced particles have energy 
larger than the cut-off scale due to the existence of the large non-minimal 
coupling~\cite{DeCross:2015uza,Ema:2016dny,Sfakianakis:2018lzf}. 
Even though a detailed study of the reheating in 2HDM inflation 
is not the scope of the current paper, it is reasonable to think that there may be a similar
issue in the 2HDM inflation without the $ R^{2} $ term; this is because 
the violent preheating is a generic feature of large non-minimal coupling. 
(However, see also Ref.~\cite{Hamada:2020kuy}) A more complete discussion of
the unitarity violation of 2HDM inflaton will be further studied elsewhere. 
Moreover, we remark that regardless of the unitarity violation, if one wants to 
have a theory valid up to Planck scale for entire field range, $R^{2}$-2HDM
inflation perhaps can be considered as a UV completion of the model as well.

One key implications of $R^2$-Higgs inflation in the g2HDM is quasi-degenerate mass spectrum for $\mathsf{H}$, $A$ and $\mathsf{H}^\pm$.
Without the confirmation of such quasi-degeneracy, a discovery of heavy Higgs bosons may not be sufficient to 
make a connection to the inflationary scenario.
Depending on the magnitude of the additional Yukawa couplings $\rho^U_{tt}$, $\rho^U_{tc}$, $\rho^D_{bb}$  etc.
such mass reconstruction may be partially possible at the LHC in certain scenarios, however, one may need future electron-positron collider
such as ILC or FCCee. 


\vskip0.2cm
\noindent{\bf Acknowledgments.--} \
The work of SML was supported in part by the National Research Foundation of Korea (NRF)
grant funded by the Korea government (MOE) (No. 2020R1A6A3A13076216). SML is also supported by the 
Hyundai Motor Chung Mong-Koo Foundation Scholarship. The work of TM is supported by a Postdoctoral 
Research Fellowship from the Alexander von Humboldt Foundation. 
The work of KO is in part supported by KAKENHI Grant Nos.~19H01899 and 21H01107.
The work of TT is supported in part by JSPS KAKENHI Grant Numbers 17H01131, 19K03874 and MEXT KAKENHI Grant Number 19H05110.

\appendix
\section{Field space metric and Christoffel symbols}\label{fieldchris}

The nonvanishing Christoffel symbols (with $M_P=1$) are
\begin{align}
%
\Gamma^{\phi_1}_{\phi_2 \phi_2}&= \Gamma^{\phi_1}_{\phi_3 \phi_3}=  \Gamma^{\phi_1}_
{\phi_4 \phi_4}= \frac{e^{-\sqrt{\frac{2}{3}} \varphi }}{\sqrt{6}}\nn\\
\Gamma^{\phi_2}_{\phi_1 \phi_2} &=\Gamma^{\phi_2}_{\phi_1 \phi_2}=-\frac{1}{\sqrt{6}},\nn\\
\Gamma^{\phi_3}_{\phi_1 \phi_3} &= \Gamma^{\phi_3}_{\phi_3 \phi_1}=-\frac{1}{\sqrt{6}},\nn\\
\Gamma^{\phi_4}_{\phi_1 \phi_4} &= \Gamma^{\phi_4}_{\phi_4 \phi_1}=-\frac{1}{\sqrt{6}}.
\end{align}

\section{The approximate initial conditions}\label{apprxini}
Let us perform following field redefinition~\cite{Gong:2012ri}:
\begin{align}
 &\rho= \frac{\sqrt{\rho_2^2 + \rho_3^2}}{\rho_1},~\tau= \frac{s}{\rho_1^2}, c_\chi = \frac{\rho_2}{\sqrt{\rho_2^2 + \rho_3^2}}\nn\\
 &~\varphi= \sqrt{\frac{3}{2}} M_P \ln\left(F^2\right)\label{redef}
\end{align}
where we have used shorthand notation $\cos\chi = c_\chi$.
The conformal factor becomes
\begin{align}
\frac{1}{F^2} = \frac{M_P^2\bigg(1- e^{-\sqrt{\frac{2}{3}}\frac{\varphi}{M_P}}\bigg)}{\xi_{11} \rho_1^2 +\xi_R s}.
\end{align}
The potential $V_E$ in Eq.~\eqref{pot} can now be expressed in terms of $(\varphi,\rho,\chi,\tau)$ as
\begin{align}
 V_E(\varphi,\rho,\tau,\chi) &=\frac{ M_P^4 \left(\eta_{\rm{eff}}+ 2 \xi_R \tau^2\right)}{8(\xi_{11}+\xi_R \tau)^2} \nn\\
 &\times \bigg(1- e^{-\sqrt{\frac{2}{3}}\frac{\varphi}{M_P}}\bigg)^2\label{eq:repot}.
\end{align}
with 
\begin{align}
 \eta_{\rm{eff}} &= \tilde{\eta_1} +\tilde{\eta_2} \rho^4+ 2 \rho^2
 \big(\tilde{\eta_3} + \tilde{\eta_4 }+ (2 c_\chi^2 -1) \tilde{\eta_5}\big)\nn\\
 & \qquad+ 4\cchi  \rho\left(\tilde{\eta_6} +\tilde{\eta_7}\rho^2\right).
\end{align}
The potential in Eq.~\eqref{eq:repot} is now in the single field attractor form with $\varphi$ playing
the role of the inflaton once it is minimized with respect to $\rho,\tau$ and $c_\chi$. Here for sake of simplicity we 
minimize first $\eta_{\rm{eff}}$ with respect to $\rho$ and $\chi$.
This is essentially minimizing the potential $V$ in the $\rho$ and $c_\chi$ direction as discussed
in the context of Higgs inflation in 2HDM in Ref.~\cite{Modak:2020fij}. We follow the same numerical minimization
procedure as in Ref.~\cite{Modak:2020fij}.
The $\eta_{\rm{eff}}$ has a extremum at 
$(\rho_0,c_{\chi_0})$, which is found by solving $\partial  V/\partial  \rho=0$ 
and $\partial  V/\partial  c_\chi=0$ simultaneously. The extremum is considered a 
minimum if both the determinant and trace of the covariant matrix 
$X_{ij}=\partial ^2 V/\partial  x_i \partial x_j$ (with $x_{i,j}= \rho~\mbox{and}~ c_\chi$),
calculated at the minima $(\rho_0,c_{\chi_0})$, are $>0$.
We find the $\eta^{\rm{min}}_{\rm{eff}}$ as
\begin{align}
 \eta^{\rm{min}}_{\rm{eff}}=&\tilde{\eta_1} +\tilde{\eta_2 }\rho_0^4+ 2 \rho_0^2
\left(\tilde{\eta_3} + \tilde{\eta_4} +  (2 c_{\chi_0}^2 -1) \tilde{\eta_5}\right) \nn\\
&+4 c_{\chi_0}  \rho_0\left(\tilde{\eta_6 }+\tilde{\eta_7} \rho_0^2\right) .
\end{align}
One can now insert $\eta^{\rm{min}}_{\rm{eff}}$ in Eq.~\eqref{eq:repot} and minimize
with respect to $\tau$ where the minimum is found as
\begin{align}
 \tau_0 = \eta^{\rm{min}}_{\rm{eff}} /(2\xi_{11}).
\end{align}
Substituting $\tau_0$ we find
\begin{align}
 V_E= \frac{M_P^4}{4}\frac{1}{(\frac{2 \xi_{11}^2}{\eta^{\rm{min}}_{\rm{eff}}}+\xi_R)}
 \bigg(1- e^{-\sqrt{\frac{2}{3}}\frac{\varphi}{M_P}}\bigg)^2\label{eq:repot2}.
\end{align}
We can now utilize Eq.~\eqref{eq:repot2} to find the $\varphi$ value that would satisfy 
the Planck 2018 measurements once the kinetic terms are canonically
normalized. We do not perform slow roll approximation, however, follow the covariant formalism and solve background field equations
Eq.~\eqref{eq:bkg} with the initial conditions of $\varphi$, $\rho_1$, $\rho_2$ and $\rho_3$ being
simply translated from these minimized values of $\rho$ ,$c_{\chi}$ and $\tau$ and $\varphi$ via Eq.~\eqref{redef}.
Here we stress the all four fields $\varphi$, $\rho_1$, $\rho_2$ and $\rho_3$ start at the top of the ridge with these initial
conditions but they quickly settles to the trajectories such that $\varphi$ essentially plays the role of inflaton.

\section{Valley Approximations} \label{Appendix:Valley}
When there is a well-defined trajectory of the inflaton with valley shaped potential, 
we have single field-like behavior and $ \rho_{i} (i=1,2,3)$ fields can be represented
as a function of $ \varphi $. In this Appendix, we present analytic understanding of these approximations.

The potential in the Einstein frame is given by
\begin{align}
	& V_E(\varphi,\rho_1,\rho_2,\rho_3)= \frac{1}{8}e^{-2\sqrt{\frac{2}{3}}\frac{\varphi}{M_P}}\bigg[ V(\rho_1,\rho_2,\rho_3) \nn\\
	&\qquad+2 \frac{M_P^4}{\xi_R}\bigg(e^{\sqrt{\frac{2}{3}}\frac{\varphi}{M_P}} - 1-\frac{\xi_{11}}{M_P^2}\rho_1^2\bigg)^2\bigg] , \label{pot:inf}
\end{align}
with
\begin{align}
	V(\rho_1,\rho_2,\rho_3) &= 
	\eta_1 \rho_1^4 +  \eta_2 \left(\rho_2^2+\rho_3^2\right)^2 +2 \eta_5\left(\rho_2^2-\rho_3^2\right) \rho_1^2+
	\nn\\
	& 2
	\left( \eta_3+  \eta_4\right) 
	\left(\rho_2^2+\rho_3^2\right) \rho_1^2+4 \rho_2 \rho_1\nn\\
	&\left\{  \eta_6 \rho_1^2 + \eta_7 \left(\rho_2^2+\rho_3^2\right) \right\} 
\end{align}
where we did not explicitly put tildes for $ \eta $s. From this, we have the following set of equations for the valley:
\begin{align}
	 \frac{\partial V_{E}}{\partial \rho_{1}}& / \left(\frac{1}{2} e^{-2\sqrt{\frac{2}{3}} \varphi}\right) =  \eta_{1} \rho_{1}^{3} + 
	2
	\frac{\xi_{11}}{\xi_{R}} \rho_{1} \left(1 - e^{\sqrt{\frac{2}{3}} \varphi} + \xi_{11} \rho_{1}^{2}\right)
	+ \nn\\ &  3 \eta_{6}\rho_{1}^{2} \rho_{2} + \eta_{5} \rho_{1} (\rho_{2} - \rho_{3}) (\rho_{2} + \rho_{3}) + 	\nn\\ 
	& (\eta_{3} + \eta_{4}) \rho_{1} (\rho_{2}^{2} + \rho_{3}^{2}) +\eta_{7} \rho_{2} (\rho_{2}^{2} + \rho_{3}^{2}) = 0, \label{Eq:1st} \\
	\frac{\partial V_{E}}{\partial \rho_{2}}& / \left(\frac{1}{2} e^{-2\sqrt{\frac{2}{3}} \varphi}\right)  = 
	\eta_{6} \rho_{1}^{3} + \rho_{2} \big[ (\eta_{3} + \eta_{4} + \eta_{5}) \rho_{1}^{2} + \nn\\
	& 3 \eta_{7} \rho_{1} \rho_{2} + \eta_{2} \rho_{2}^{2} \big]
	+ (\eta_{7} \rho_{1} +\eta_{2} \rho_{2}) \rho_{3}^{2} = 0, \label{Eq:2nd} \\
	\frac{\partial V_{E}}{\partial \rho_{3}}& / \left(\frac{1}{2} e^{-2\sqrt{\frac{2}{3}} \varphi}\right)  = 
	\rho_{3} \big[(\eta_{3}+\eta_{4}- \eta_{5}) \rho_{1}^{2} + 2 \eta_{7}\rho_{1} \rho_{2} +\nn\\
	&\qquad\qquad\qquad \qquad\qquad \eta_{2}\rho_{2}^{2} + \eta_{2} \rho_{3}^{2} \big] = 0. \label{Eq:3rd}
\end{align}
From the last equation Eq.~\eqref{Eq:3rd}, we have $ \rho_{3} = 0 $.

Then Eq.~\eqref{Eq:2nd} reduces to
\begin{align}
	\eta_{2} x^{3} + 3 \eta_{7} x^{2} + (\eta_{3} + \eta_{4} + \eta_{5}) x + \eta_{6} = 0 \,,
\end{align}
where $ x \equiv \frac{\rho_{2}}{\rho_{1}} $. For our parameters, we have one real solution, 
which is denoted by $ x=C $. Then we have $ \rho_{2} = C\rho_{1}  $.
Finally, by having $ \rho_{3} = 0 $ and $ \rho_{2} = C \rho_{1} $, Eq.~\eqref{Eq:1st} gives
\begin{align}
\rho_{1} &=	D \sqrt{e^{\sqrt{\frac{2}{3}} \varphi}-1}~~~~\mbox{with}, \\ 
& D \equiv \frac{\sqrt{2\xi_{11}}}{\sqrt{
		\xi_{R} 
		(C^{3} \eta_{7} 
		+C^{2}(\eta_{3}+\eta_{4}+\eta_{5})
		+3C \eta_{6}+ \eta_{1})
		+2 \xi_{11}^2}}\nn.
\end{align}



\end{document}